\documentclass[12pt, draftclsnofoot, onecolumn]{IEEEtran}
\normalsize
\IEEEoverridecommandlockouts
\usepackage{graphicx, amsmath, amssymb, subcaption, url, cite, array}
\usepackage[ruled, linesnumbered]{algorithm2e} 
\usepackage{algorithm2e, setspace}
\usepackage[font={small}]{caption}
\usepackage[hidelinks]{hyperref}
\graphicspath{ {./figures/} }
\DeclareMathOperator*{\argmax}{\arg\!\max}
\DeclareMathOperator*{\argmin}{\arg\!\min}

\newcommand*\Eval[3]{\left.#1\right\rvert_{#2}^{#3}}
\newtheorem{proposition}{\textbf{Proposition}}

\begin{document}

\title{\huge Joint Power Allocation and Rate Control for Rate Splitting Multiple Access Networks with Covert Communications}

\author{Nguyen Quang Hieu, Dinh Thai Hoang, Dusit Niyato, Diep N. Nguyen, \\Dong In Kim, and Abbas Jamalipour
\thanks{N. Q. Hieu, D. T. Hoang, and D. N. Nguyen are with the School of Electical and Data Engineerng, University of Technology Sydney, Sydney, NSW 2007 (email: hieu.nguyen-1@student.uts.edu.au, \{hoang.dinh, diep.nguyen\}@uts.edu.au).}
\thanks{D. Niyato is with the School of Computer Science and Engineering, Nanyang Technological University, Sinapore  (e-mail: dniyato@ntu.edu.sg). }
\thanks{D. I. Kim is with the Department of Electrical and Computer Engineering, Sungkyunkwan University (SKKU), Suwon 16419, South Korea (e-mail: dikim@skku.ac.kr).}
\thanks{A. Jamalipour is with the School of Electrical and Information Engineering, The University of Sydney, Australia, NSW 2006 (e-mail: a.jamalipour@ieee.org).}
}

\maketitle
\begin{abstract}
Rate Splitting Multiple Access (RSMA) has recently emerged as a promising technique to enhance the transmission rate for multiple access networks. Unlike conventional multiple access schemes, RSMA requires splitting and transmitting messages at different rates. The joint optimization of the power allocation and rate control at the transmitter is challenging given the uncertainty and dynamics of the environment.
Furthermore, securing transmissions in RSMA networks is a crucial problem because the messages transmitted can be easily exposed to adversaries. This work first proposes a stochastic optimization framework that allows the transmitter to adaptively adjust its power and transmission rates allocated to users, and thereby maximizing the sum-rate and fairness of the system under the presence of an adversary. We then develop a highly effective learning algorithm that can help the transmitter to find the optimal policy without requiring complete information about the environment in advance. Extensive simulations show that our proposed scheme can achieve positive covert transmission rates in the finite blocklength regime and non-saturating rates at high SNR values. More significantly, our achievable covert rate can be increased at high SNR values (i.e., 20 dB to 40 dB), compared with saturating rates of a conventional multiple access scheme.
\end{abstract}

\begin{IEEEkeywords}
Rate splitting multiple access, covert communications, deep reinforcement learning, power allocation, rate control.
\end{IEEEkeywords}
\IEEEpeerreviewmaketitle

\section{Introduction}
Recent years have witnessed the growing interest in six-generation (6G) networks from both academia and industry. It is envisioned that 6G will enable the Internet-of-Things (IoT) in which a massive number of devices can communicate via wireless environments. To accommodate such a growing demand of connections in future wireless networks, a modern multiple access scheme with high efficiency and flexibility is an urgent need. Rate Splitting Multiple Access (RSMA) has recently emerged as a novel communication technique that can flexibly and efficiently manage interference and thus increase the overall performance for the downlink of the wireless systems~\cite{dizdar2020, dizdar2020-2}. In RSMA, each message at the transmitter is first split into two parts, i.e., a common part and private part. The common parts of all the messages are combined into a single common message. The common message is then encoded using a shared codebook. The private messages are independently encoded for the respective users. After receiving these messages, each user decodes the common and private messages with the Successive Interference Cancellation (SIC) to obtain its original message. By partially decoding and partially treating interference as noise, RSMA can enhance spectral efficiency, energy efficiency, and security of multiple access systems~\cite{mao2018, li2020, joudeh2016-2, joudeh2016, mao2018-2, mao2021}. Thanks to its outstanding features, RSMA can tackle many emerging problems in 6G and gain enormous attention from both industry and academia~\cite{dizdar2020-2}.

\subsection{Challenges of RSMA}
Although possessing the advantages, the RSMA scheme is facing two main challenges. Unlike conventional multiple access schemes, e.g., Spatial Division Multiple Access (SDMA) which treats interference as noise or Non-Orthogonal Multiple Access (NOMA) which successively removes multi-user interference during the decoding process, RSMA partially decodes the messages and partially treat multi-user interference as noise. The RSMA transmitter hence needs to jointly optimize the power allocation and rate control for different messages to maximize the energy and spectral efficiency for the whole system. Aiming to address such problems, several research works have been proposed in the literature. In~\cite{piovano2016}, the authors consider a rate splitting approach with heterogeneous Channel State Information at the Transmitter (CSIT). Two groups of CSIT qualities are considered and the transmitter is assumed to have either partial CSIT or no-CSIT. For the no-CSIT scenario, the users decode their messages by treating interference as noise without using a rate splitting strategy. In contrast, for the partial CSIT scenarion, a rate splitting strategy is applied and a fraction of total power is allocated equally among private symbols and common symbols. Simulation results show that with the proposed power allocation scheme, the sum-rate of users in the group with the rate splitting strategy can gain significant improvement compared to those in the group without using any rate splitting strategy. Similarly, in~\cite{joudeh2016}, the authors study a rate splitting strategy for a multi-group of users in a large-scale RSMA system. A more complex precoding design for power and rate allocated to users is proposed to find the maximum sum-rate of the system. The simulation results reveal that a precoding design with rate splitting can benefit sum-rate of the system. However, the authors in~\cite{joudeh2016} only consider a perfect CSIT scenario in which the BS is assumed to know exact information of the channel and the channel is assumed to be fixed. In order to relax these assumptions, the authors in~\cite{joudeh2016-2} and~\cite{joudeh2016-3} consider a similar system, but under imperfect CSIT. In this case, a stochastic optimization formulation is proposed to deal with the uncertainty of the channel with imperfect CSIT. In~\cite{joudeh2016-3}, the authors show that with rate splitting under imperfect CSIT, the sum-rate of the system can achieve non-saturating rates at high SNR values compared to saturating rates of a conventional scheme, i.e., SDMA. In~\cite{joudeh2016-2}, the authors further investigate the impacts of different error models on the system performance. Numerical results reveal that in addition to the expected sum-rate gains, the benefits of rate splitting also include relaxed CSIT quality requirements and enhanced achievable rate regions compared with a conventional transmission scheme. A comprehensive analysis of RSMA performance is studied in~\cite{mao2018}. In this work, through many simulations and performance analysis, they show that the rate splitting techniques are able to softly bridge the two extremes of fully treating interference as noise and fully decoding interference. Thus, in comparison with conventional multiple access approaches, e.g., SDMA and NOMA, RSMA can gain significant rate enhancement. Although the aforementioned works propose solutions to improve system performance for RSMA networks, the channel state distribution (or channel state matrix) is always assumed to be known by the transmitter. In addition, optimization methods in these works also introduce additional variables, e.g., equalizers and weights, that are highly correlated with the channel state. Thus, unavailability or drastic changes of the channel state information, e.g., due to mobility of users of link’s failures~\cite{guo2017}, can result in significant degradation of these algorithms. Therefore, a more flexible framework that not only deals with the dynamics of the environment but also efficiently manages power allocation and rate control for RSMA without requiring complete or partial information, e.g., posterior distributions, of the channel state, is in an urgent need.

The second challenge that RSMA is facing is security. Although a new data rate region can be achieved with RSMA, investigation on RSMA's security is still in its early stage. Several works, such as~\cite{li2020, fu2020} and~\cite{xia2022}, are proposed to address the eavesdropping issues in RSMA networks. In particular, the authors in~\cite{li2020} propose a cooperative RSMA scheme to enhance the secrecy sum-rate of the system in which the common messages can be used not only as desired messages but also artificial noise. In this case, it is shown that the proposed cooperative secure rate splitting scheme outperforms conventional SDMA and NOMA schemes in terms of secrecy sum-rate. However, \cite{li2020} only considers the perfect CSIT scenario in which the transmitter has complete information of the channel state. In order to address the challenges caused by imperfect CSIT, the authors in~\cite{fu2020} investigate the impacts of imperfect CSIT on the secrecy rate of the system. Specifically, to deal with imperfect CSIT, a worst-case uncertainty channel model is taken into consideration with the goal to mitigate simultaneously inter-user interference and maximize the secrecy sum-rate. The simulation results show the robustness of the proposed solution against the imperfect CSIT of RSMA, and the secure transmission is also guaranteed. Furthermore, in comparison with NOMA, RSMA shows significant secrecy rate enhancement. In~\cite{xia2022}, a secure beamforming design is also proposed to maximize the weighted sum-rate under user's secrecy rate requirement. Unlike~\cite{li2020} and~\cite{fu2020}, the authors in~\cite{xia2022} consider the presence of an internal eavesdropper, i.e., an illegitimate user, that not only receives its messages but also wiretaps messages intended for other legitimate users. To deal with this internal eavesdropper, all user's secrecy rate constraints are taken into consideration. The simulation results suggest that RSMA can outperform the baseline scheme in terms of weighted sum-rate. 

All of the above works and others in literature only focus on dealing with the passive eavesdroppers, i.e., the eavesdroppers try to listen passively to the communication channel to derive the original message. To deal with such passive eavesdroppers, the transmitter can adaptively select different transmission rates~\cite{xia2022} or utilize artificial noise~\cite{li2020} to confuse the eavesdropper, and thereby minimizing the information disclosure. However, in these cases, the eavesdropper can still detect and receive the signals from the transmitter, and thus it can still decode the information if it has a more powerful hardware computation, e.g., through employing cooperative processing with other eavesdroppers, or better antennas gains compared with those of the  transmitter~\cite{trappe2015, he2010}. The passive eavesdropper  scenario cannot address the problem in which the eavesdropper is able to manipulate or control the environment~\cite{trappe2015}. For example, by manipulating the environment, the adversary can bias the resulting bits in the key establishment process~\cite{jana2009}. 
In applications requiring a high security protection, e.g., military and IoT healthcare applications, leaking a small amount of data can result in a break of the whole system and/or cause effects to the users. Therefore, in this work, to further prevent potential information leakage, we focus on a more challenging adversary model in which we need to control the power and transmission rates allocated to users, so that the adversary is unable to detect transmissions on the channels. In this way, the possibility of leaking information can be minimized.

\subsection{Contributions and Organization}
In this work, we aim to develop a novel framework that addresses the aforementioned problems. In particular, we consider a scenario in which an adversary, i.e., a warden, is present in the communication range of a Base Station (BS), i.e., the transmitter, and multiple mobile users, i.e., the receivers. The warden is assumed to be able to observe constantly the channel with a radiometer and interrupt the channel if it detects transmissions on the channel~\cite{yan2019, yan2018}. Thus, it is challenging for the BS to allocate jointly power and control transmission rates for all the messages while hiding these messages from the warden. To minimize the probability of being detected by the warden, a possible policy for the BS is to decrease the power allocated to the messages, so that the warden can be confused the transmitted signals with noise. However, an inappropriate implementation can result in zero data rates at the receivers~\cite{yan2019}. To maximize the transmission rate of the system and at the same time guarantee non-zero rate at each user, we formulate the problem of the BS as a max-min fairness problem. In particular, the BS aims to maximize the expected minimum rate (min-rate) of the system under the uncertainty of the environment. Furthermore, we consider a covert constraint that is derived from the theory of covert communications (i.e., low probability of detection communications)~\cite{ tao2020, jiang2020, forouzesh2021}. In this way, our proposed framework can help the BS secretly communicate with the legitimate mobile users with a small probability of being detected by the warden.

To find the optimal solution for the optimization formulation above, we develop a learning algorithm based on Proximal Policy Optimization (PPO) ~\cite{schulman2017}. By leveraging recent advances of deep reinforcement learning techniques~\cite{luong2019, kiran2021, sutton2018}, our proposed algorithm can effectively find the optimal policy for the BS without requiring complete information of the channel in advance. Specifically, our proposed algorithm takes the feedback from users via the uplink as inputs of the deep neural networks and then outputs the corresponding joint power allocation and rate control policy for the BS. This procedure is similar to the conventional optimization-based schemes in the view of implementation and resource. The differences in our proposed algorithm are twofold. First, our proposed algorithm is a model-free deep reinforcement learning algorithm which does not require a complete model of the environment, i.e., the channel state matrix or channel distribution, in advance. Second, the policy obtained by our proposed algorithm can be adaptively adjusted in cases the channel dynamics change over time, e.g., time-varying channel. Thus, our proposed algorithm is expected to show more flexibility and robustness against the uncertainty and dynamics of the wireless environment.
With the obtained optimal policy, we then show that our proposed method can also achieve covert communications between the BS and mobile users by dynamically adjusting power and transmission rates allocated to the messages. 

Here, we note that our previous work presented in~\cite{hieu2021} only focuses on power allocation problem without considering the security impact on the RSMA system.  Furthermore, in this current work, we further investigate the rate and security performance of RSMA in the finite blocklength (FBL) regime where the achievable covert rate is limited and no longer follows the Shannon capacity (i.e., infinite blocklength regime). Thus, our framework can be applicable for a wide range of applications which include transmission between IoT devices where the transmitted data is expected to be sporadic and the number of channel uses is limited. In short, our main contributions are as follows:

\begin{itemize}
\item We develop a novel stochastic optimization framework to achieve the max-min fairness of the considered RSMA network with the covert constraint. This framework enables the BS to make optimal decisions to maximize the expected min-rate under the covert requirement as well as the dynamics and uncertainty of surrounding environment. To the best of our knowledge, this is the first work that considers covert communications for RSMA. Therefore, our proposed framework is a promising solution for secure and reliable, high data transmission rate applications.
\item We propose a highly effective learning algorithm to make the best decisions for the BS. This learning algorithm enables the BS to quickly find the optimal policy through feedback from the users by leveraging advantages of both deep learning and reinforcement learning techniques. Furthermore, our proposed learning algorithm can effectively handle the continuous action and state spaces for the BS through using the PPO technique. 
\item We conduct intensive simulations to evaluate the efficiency of the proposed framework and reveal insightful information. Specifically, the simulation results show that the positive covert rate is achievable with RSMA in the finite blocklength regime where the achievable covert rate is limited and no longer follows the Shannon capacity (i.e., infinite blocklength regime). More interestingly, with high values of transmission power (i.e., 20 dB to 40 dB), our achievable covert rate can be increased while the achievable covert rate of a baseline multiple access scheme, i.e., SDMA, is saturated. Thus, beyond conventional wireless networks, our framework can be applicable for IoT networks in which the data transmitted between devices is expected to be sporadic and with a relatively small quantity of information .
\end{itemize}

The rest of our paper is organized as follows. Our system model is described in Section~\ref{sec:system-model}. Then, we formulate the stochastic optimization problem for the covert-aided RSMA networks in Section~\ref{sec:problem-formulation}. We provide details of our proposed learning algorithm to maximize the covert rate of the system in Section~\ref{sec:ppo-algo}. After that, our simulation results are discussed in Section~\ref{sec:performance-evaluation}, and Section~\ref{sec:conclusion} concludes the paper.

\section{System Model}
\label{sec:system-model}
\begin{figure}
\centering
\includegraphics[width=0.9\linewidth]{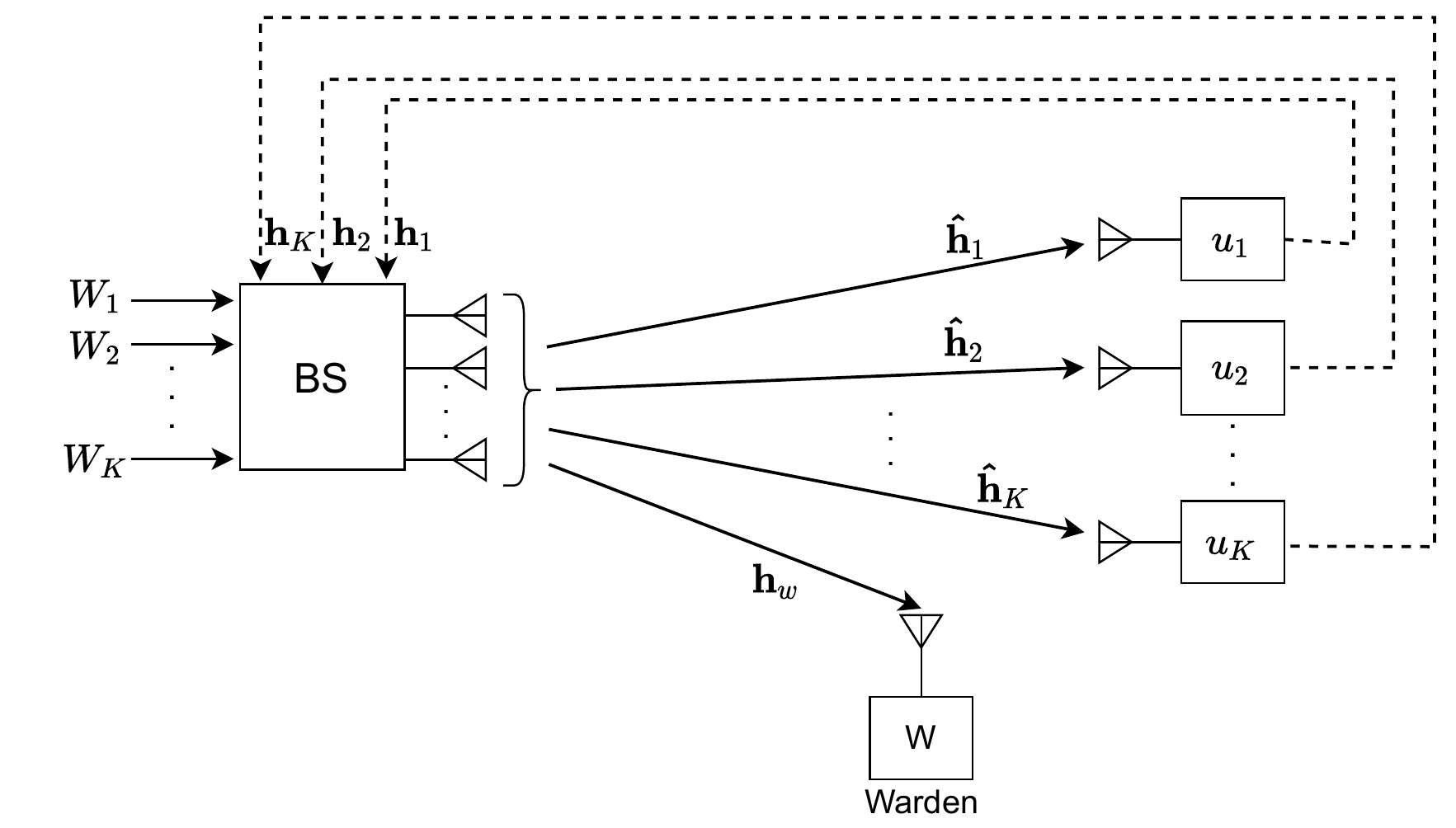}
\caption{Covert-aided RSMA system model.}
\label{fig:system-model}
\end{figure}

We consider a system that consists of one $M$-antenna BS, a set $\mathcal{K} = \{1, 2, \ldots, K\}$ of $K$ single-antenna legitimate users ($M \geq K$), and a warden as illustrated in Fig.~\ref{fig:system-model}. The warden has ability to interrupt the channel if it detects any transmissions from the BS.
The BS wants to transmit information to the users with a minimum probability of being detected by the warden~\cite{bash2013}. The BS has a set of messages $\mathbf{W} = \{W_1, \ldots, W_k, \ldots, W_K\}$ to be transmitted to the users. The message intended for user $u_k$, denoted as $W_k$, is split into a common part $W_k^c$ and a private part $W_k^p$ ($\forall k \in \mathcal{K}$), with the lengths of $l_k^c$ and $l_k^p$, respectively. The common parts of all $K$ messages are combined into a single common message $W^c$. The single common message $W^c$ and $K$ private messages $W_k^p$ are independently encoded into streams $s_c$ (i.e., common stream), $s_1, s_2, \ldots, s_K$ (i.e., private streams), respectively. The transmitted signal of the BS is thus defined as follows:
\begin{equation}
\mathbf{x} = \mathbf{p}_c s_c + \sum_{k=1}^{K} \mathbf{p}_k s_k,
\end{equation} 
where $\mathbf{p}_c$ and $\mathbf{p}_k$ are the beamforming vectors for the common and private stream $s_c$ and $s_k$, respectively. Let $\mathbf{h}_k \in \mathbb{C}^{M\times 1}$ denote the estimated channel between the BS and user $u_k$, $\mathbf{h}_w \in \mathbb{C}^{M\times 1}$ denote the channel between the BS and the warden. The received signal at user $u_k$ is calculated as follows:
\begin{equation}
y_k = \mathbf{h}_k^H \mathbf{x} + n_k,
\end{equation}
where $n_k \sim \mathcal{CN}(0, \sigma_{n,k}^2)$ is the Additive White Gaussian Noise (AWGN) at the receiver. Note that the estimated channel at the BS is obtained from feedback of the users which contains estimation errors, i.e., $\mathbf{h}_k = \mathbf{\hat{h}}_k + \mathbf{\tilde{h}}_k$, where $\mathbf{\hat{h}}_k$ is the actual channel state and $\mathbf{\tilde{h}}_k$ is the estimation error.
The SINRs of the common and private messages, denoted as $\gamma_k^c$ and $\gamma_k^p$, respectively, can be calculated as follows:
\begin{equation}
\begin{aligned}
\label{eq:sinr}
 & \gamma_k^c(\mathbf{P})= \frac{|\mathbf{h}_k^H \mathbf{p}_c|^2}{\sum_{j=1}^{K}|\mathbf{h}_k \mathbf{p}_k|^2 + 1},  \\
 & \gamma_k^p(\mathbf{P}) = \frac{|\mathbf{h}_k^H \mathbf{p}_k|^2}{\sum_{j \neq k}|\mathbf{h}_k \mathbf{p}_k|^2 + 1},
\end{aligned}
\end{equation}
where $\mathbf{P} = [\mathbf{p}_c, \mathbf{p}_1, \ldots, \mathbf{p}_K]$ is the beamformer of the BS. The transmission power at the BS is constrained by  $\text{tr}(\mathbf{P} \mathbf{P}^H) \leq P_t$. 

In the FBL regime, the achievable covert rate of the private message $W_k^p$ at user $u_k$ is calculated as follows~\cite{sun2018, yan2017, shu2019}:
\begin{equation}
\begin{aligned}
R_k^p(\mathbf{P}, l_k^p) \approx \log_2(1+\gamma_k^p) - \sqrt{\frac{\gamma_k^p(\gamma_k^p+2)}{l_k^p (\gamma_k^p+1)^2}} \frac{Q^{-1}(\delta_k)}{\ln 2}  + \frac{\log_2 l_k^p}{2l_k^p},
\end{aligned}
\end{equation}
where $l_k^p$ is the length of message $W_k^p$. $\delta_k$ is the decoding error probability at user $u_k$, and $Q^{-1}(\cdot)$ is the inverse Q-function of $Q(x) = \int_{x}^{\infty} \exp(\frac{-t}{2})dt$~\cite{sun2018}.  To guarantee that the common message $W^c$ can be correctly decoded by all the users, the achievable covert rate of the common message is calculated by~\cite{sun2018, yan2017}:
\begin{equation}
\begin{aligned}
R_k^c(\mathbf{P}, l_k^c) \approx  \min_{k \in \mathcal{K}}\Big(\log_2(1+\gamma_k^c) - \sqrt{\frac{\gamma_k^c(\gamma_k^c+2)}{l_k^c(\gamma_k^c+1)^2}} \frac{Q^{-1}(\delta_k)}{\ln 2} + \frac{\log_2 l_k^c}{2 l_k^c} \Big).
\end{aligned}
\end{equation}
Since $R_k^c$ is shared between users such that $C_k$ is the user $u_k$'s portion of the common rate $R_c$ with $\sum_{k=1}^{K}C_k \leq R_k^c$. The total achievable rate of the user $u_k$ is then defined by~\cite{mao2018}: 
\begin{equation}
R_k^{tot} = C_k + R_k^p.
\end{equation}
As a result, the covert sum-rate of the BS is defined by a sum of $R_k^{tot}$ over $K$ users, i.e., $R_s =  \sum_{k=1}^K R_k^{tot}$.

With the presence of noise, the warden needs to make a binary decision, i.e., (i) the BS is transmitting or (ii) the BS is not transmitting, based on its observations~\cite{bash2013}. For this, the warden distinguishes two hypotheses $\mathcal{H}_0$ and $\mathcal{H}_1$, where $\mathcal{H}_0$ denotes the null hypothesis, i.e.,  the BS is not transmitting, and $\mathcal{H}_1$ denotes the alternative hypothesis, i.e.,  the BS is transmitting. In particular, the two hypotheses are defined as follows:
\begin{equation}
\begin{cases}
\mathcal{H}_0\ : & \mathbf{y}_w = \mathbf{z}_w \\ 
\mathcal{H}_1\ : & \mathbf{y}_w = \mathbf{h}_w^H \mathbf{x} + \mathbf{z}_w,
\end{cases}
\label{eq:hypothesis-test}
\end{equation}
where $\mathbf{y}_w$ and $\mathbf{z}_w$ are the received signal and noise signal at the warden, respectively. 
$\mathbf{x}$ is the transmitted signal from the BS. It is noted that the warden does not know the codebook of transmitted signals and the hypothesis test of the warden can be performed as follows. First, the warden collects a row vector of independent readings $\mathbf{y}_w$ from his channel to the BS. Then the warden generates the test statistic on the collected vector.
The goal of the warden is to minimize the error detection rate, which is given by:
\begin{equation}
\xi = P_F + P_M,
\end{equation}
where $P_F = \text{Pr}(\mathcal{D}_1|\mathcal{H}_0)$ is the false alarm probability and $P_M = \text{Pr}(\mathcal{D}_0|\mathcal{H}_1)$ is the miss detection probability. $\mathcal{D}_1$ and $\mathcal{D}_0$ are the binary decisions of the warden that infer whether the BS is transmitting or not, respectively. We can derive the lower bound of $\xi$ as follows~\cite{yan2017}:
\begin{equation}
\label{eq:lower-bound-error}
\xi \geq 1 - \sqrt{\frac{1}{2}\mathcal{D}(\mathbb{P}_0||\mathbb{P}_1)},
\end{equation}
where $\mathbb{P}_0$ and $\mathbb{P}_1$ are the probability distributions of the observations when the BS transmits (i.e., $\mathcal{H}_1$ is true) or when the BS does not transmit (i.e., $\mathcal{H}_0$ is true), respectively. 
$\mathcal{D}(\mathbb{P}_0||\mathbb{P}_1)$ is the relative entropy  between two probability distributions $\mathbb{P}_0$ and $\mathbb{P}_1$.
\begin{proposition}
\label{prop:kl-div}
 The relative entropy (Kullback–Leibler divergence) between two probability distributions $\mathbb{P}_0$ and $\mathbb{P}_1$, denoted by  $\mathcal{D}(\mathbb{P}_0||\mathbb{P}_1)$, can be calculated as follows:
\begin{equation}
\label{eq:KL-divergence}
\mathcal{D}(\mathbb{P}_0 | \mathbb{P}_1) = \ln\left(\sqrt{g_w P_t + \sigma_w^2}\right) - \ln\left(\sigma_w\right) + \frac{\sigma_w^2}{2\left(g_w P_t + \sigma_w^2\right)} - \frac{1}{2}.
\end{equation}
\end{proposition}

Proof of Proposition \ref{prop:kl-div} can be found in Appendix~\ref{appe:kl}.

In covert communications, we normally have $\xi \geq 1 - \epsilon$ as the covertness requirement, where $\epsilon$ is an arbitrarily small value~\cite{yan2017}. Following (\ref{eq:lower-bound-error}), in this paper, we adopt $\mathcal{D}(\mathbb{P}_0||\mathbb{P}_1) \leq 2 \epsilon^2$ as the covertness requirement.

In the covert-aided systems, the achievable data rate is usually small or asymptotically approached zero~\cite{bash2013, yan2017}. To achieve the maximum rate of the system and non-zero  data rate for each user, we consider a max-min fairness problem in which the optimization problem is formulated as maximizing the expected minimum data rate (min-rate) among users~\cite{joudeh2016}. Let $\mathbf{L} = [l_1^c, l_1^p, \ldots, l_K^c, l_K^p]$ denote the message-splitting vector and $\mathbf{C} = [C_1, C_2, \ldots, C_K]$  denote the common rates for the common messages. The stochastic optimization problem of the BS is then formulated as follows:
\begin{subequations}
\label{eq:max-min-rate}
\begin{align}
\max_{\mathbf{P}, \mathbf{L}, \mathbf{C}} \quad & \min_{k \in \mathcal{K}} \bar{R}_k^{tot} \\
\textrm{s.t.} \quad & \text{tr}(\mathbf{P} \mathbf{P}^H) \leq P_t,\\
\quad & l_k^c + l_k^p = L_k, \forall k \in \mathcal{K}, \\
\quad & \sum_{k \in \mathcal{K}} C_k \leq R_c, \forall k \in \mathcal{K}, \\
\quad & \mathcal{D}(\mathbb{P}_0||\mathbb{P}_1) \leq 2\epsilon^2,\\
\quad & R_k^{tot} \geq R_k^0,
\end{align}
\end{subequations}
where $\bar{R}_k^{tot} = \mathbb{E}_{\mathbf{h}_k \in \mathbf{H}} \{ R_k^{tot}\}$ is the average rate of the system with $\mathbf{H} = \{\mathbf{h}_k | \mathbf{h}_k = \mathbf{\hat{h}}_k + \mathbf{\tilde{h}}_k; \forall k \in \mathcal{K}\}$ is the channel matrix of the system.
$R_k^0$ is the minimum rate requirement (QoS) of user $u_k$ and $L_k$ is length of the message $W_k$. The problem in (\ref{eq:max-min-rate}) can be described as follows. (\ref{eq:max-min-rate}b) and (\ref{eq:max-min-rate}c) illustrate the power constraint and packet length constraint of the BS, respectively. (\ref{eq:max-min-rate}d) is the common rate constraint. (\ref{eq:max-min-rate}e) and (\ref{eq:max-min-rate}f) are covert constraint and QoS constraint, respectively.
Optimizing (\ref{eq:max-min-rate}) is very challenging under the dynamics and uncertainty of the communication channel, i.e., the channel gain $\mathbf{h}_k$ between the BS and user $u_k$ changes over time, and the channel state is unknown to the BS. In this paper, we thus propose a deep reinforcement learning approach to obtain the optimal policy for the BS under the dynamics and uncertainty of the environment. It is noted that we only use the channel matrix $\mathbf{H}$ in the optimization problem above to illustrate the stochastic nature of the system. In the next section, we show that the optimization problem (\ref{eq:max-min-rate}) can be transformed into maximizing the expected discounted reward in the deep reinforcement learning setting without requiring any information from channel matrix $\mathbf{H}$. Details of notations used in this paper are summarized in Table~\ref{table:notations}.

\begin{table*}
\centering
\caption{Summary of notations.}
\begin{tabular}{|c|c|}
\hline \textbf{Variable} & \textbf{Definition} \\
\hline$K$ & Number of users \\
\hline$M$ & Number of antennas at the BS \\
\hline$W_k$ & Message intended to transmit to user $u_k$ \\
\hline$W_k^c, W_k^p$ & Common and private parts split from $W_k$\\
\hline$L_k$ & Length of $W_k$ (bits) \\
\hline$l_k^c, l_k^p$ & Lengths of $W_k^c$ and $W_k^p$\\
\hline $P_t$ & Transmission power of the BS  \\
\hline $\mathbf{P}$ & Transmission beamformer of the BS \\
\hline $\mathbf{L}$ & Vector of messages' lengths at the BS \\
\hline $\mathbf{C}$ & Common rate vector allocated to users  \\
\hline $\gamma_k^c (\mathbf{P}), \gamma_k^p(\mathbf{P})$ & SINRs of the common and private messages at user $u_k$ \\
\hline $\mathbf{h}_k$ & Channel between the BS and user $u_k$ \\
\hline $\mathbf{h}_w$ & Channel between the BS and warden \\
\hline $R_k^c(\mathbf{P}, l_k^c), R_k^p(\mathbf{P}, l_k^p)$ & Achievable covert rates of $W_k^c$ and  $W_k^p$\\
\hline $R_k^{tot}$ & Achievable rate of user $u_k$ \\
\hline $R_s(\mathbf{P}, L_k)$ & Achievable (covert) sum-rate \\
\hline $\epsilon$ & Covert requirement \\
\hline $\mathcal{D}(\mathbb{P}_0||\mathbb{P}_1)$ & Relative entropy between two probability distributions $\mathbb{P}_0$ and $\mathbb{P}_1$\\
\hline $R_k^0$ & Minimum rate requirement of user $u_k$ \\
\hline $\mathcal{S}, \mathcal{A}$ & State space and action space of the BS \\
\hline $s_t, a_t, r_t(s_t, a_t)$ & State, action, and reward of the BS at time step $t$ \\
\hline $p_t$ & Penalty of the BS for taking action $a_t$ \\
\hline $\Omega_{\theta}, \theta$ & Policy and policy parameter vector of the BS \\
\hline
\end{tabular}
\label{table:notations}
\end{table*}

\section{Problem Formulation}
\label{sec:problem-formulation}
\subsection{Deep Reinforcement Learning}
\label{subsec:drl-intro}
Before introducing our mathematical formulation, we first describe the fundamentals of DRL. In conventional reinforcement learning (RL) settings, an agent aims to learn an optimal policy through interacting with an environment in discrete decision time steps. At each time step $t$, the agent first observes its current state $s_t$ in a state space $\mathcal{S}$ of the system. Based on the observed state $s_t$ and current policy $\Omega$, the agent takes an action $a_t$ in the action space $\mathcal{A}$. The policy $\Omega$ can be a mapping function from a state to an action (deterministic) or a probability distribution over actions (stochastic). After taking the action $a_t$, the agent transits to a new state $s_{t+1}$ and observes an immediate reward $r_t$. The goal of the agent is to find an optimal policy that can be obtained by maximizing a discounted cumulative reward.

In conventional RL settings, the agent usually deals with a policy search problem in which the convergence time of the RL algorithm depends on the search space $\mathcal{S}$ and $\mathcal{A}$. In environments with large discrete state-action space or continuous state-action space, the optimal policy is either nearly impossible or time-consuming to find. To address this problem, RL algorithms combined with deep neural networks, namely DRL, show significant performance improvements over conventional RL algorithms~\cite{mnih2015}. In DRL algorithms, the policy $\Omega$ is defined by a probability distribution over actions, i.e., $\Omega_{\theta} = Pr\{a_t|s_t;\theta\}$, where $\theta$ is a parameter vector of the deep neural network. The parameter vector $\Omega_{\theta}$ can be trained by action-value methods, e.g., DQN~\cite{mnih2015}, or policy gradient methods, e.g., PPO~\cite{schulman2017}. Action-value methods and policy gradient methods have their advantages and drawbacks which we will discuss later in Section~\ref{subsec:optimization-formulate}. In the following, we formulate our considered problem in the DRL setting by defining state space, action space, and immediate reward function in which the BS is empowered by an intelligent DRL agent.

\subsection{DRL-based Optimization Framework}
We introduce the proposed DRL-based optimization framework for the joint power allocation and transmission rate control problem of the BS as follows. The state space of the BS is defined by:
\begin{equation}
\mathcal{S} = \Big \{ \{\mathbf{h}_k,  L_k\}; 1 \leq k \leq K \Big\},
\end{equation}
where $\mathbf{h}_k$ is the channel state feedback of the user $u_k$ to the BS. $L_k$ is the length of the message $W_k$ intended for user $u_k$. Note that the channel state feedback from the users contains estimation errors, i.e., $\mathbf{h}_k = \mathbf{\hat{h}}_k + \mathbf{\tilde{h}}_k$, where $\mathbf{\hat{h}}_k$ is the actual channel state and $\mathbf{\tilde{h}}_k$ is the estimation error. The channel of user $u_k$ is realized as
\begin{equation}
 \hat{\mathbf{h}}_k = g_k \times [1, e^{j\phi_k}, e^{j2\phi_k}, e^{j3\phi_k}],
 \label{eq:channel-realization}
\end{equation}
where $g_k \in \mathbb{R}$ and $\phi \in \mathbb{R}$ are control variables~\cite{mao2018}. The channel estimation error follows a complex Gaussian distribution, i.e., $\tilde{\mathbf{h}}_k \sim \mathcal{CN}(0, \sigma_k^2)$, where $\sigma_k^2$ is inversely proportional to the transmission power at the BS, i.e., $\sigma_k^2 = g_k P_t^{-\alpha_k} $where $\alpha_k$ is the degree of freedom (DoF) variable~\cite{mao2018}.
Note that the channel $\mathbf{h}_w$ of the warden is unknown to the BS and thus $\mathbf{h}_w$ is not included in the state space of the BS. We define the channel between the warden and the BS as follows:
\begin{equation}
\mathbf{h}_w = g_w \times [1, e^{j\phi_w}, e^{j2\phi_w}, e^{j3\phi_w}],
\label{eq:channel-warden}
\end{equation}

At each time step $t$, the BS allocates the transmission power to the users, splits the messages to common and private messages, and controls transmission rate for the messages. Thus, the action space of the BS is defined as follows:
\begin{equation}
\mathcal{A} = \{\mathbf{P}, \mathbf{L}, \mathbf{C}\}.
\end{equation}

The reward function is designed to maximize the min-rate of the BS as in (\ref{eq:max-min-rate}).  To encourage the BS to optimize the min-rate while the covert and QoS constraints of users are taken into consideration, we penalize the BS for each violated constraint. For this, the immediate reward can be defined as follows:
\begin{equation} 
r_t(s_t, a_t) = 
\begin{cases}
      \min_{k \in \mathcal{K}} R_k^{tot}, & \text{if}\ p_t = 0, \\
      0, & \text{if}\ p_t > 0.
    \end{cases}
\end{equation}
where $p_t$ is the penalty received by the BS for action $a_t$ that does not satisfy the covert constraint and QoS constraint in (\ref{eq:max-min-rate}). The penalty $p_t$ is defined as follows:
\begin{equation}
\begin{aligned}
p_t 
& = \frac{\beta_{0} \mathbf{1}\left(\mathcal{D}\left(\mathbb{P}_{0} \| \mathbb{P}_{1}\right)-2 \epsilon^{2}\right)+\sum_{k=1}^{K} \beta_{k} \mathbf{1}\left(R_{k}^{0}-R_{k}^{t o t}\right)}{\beta_{0}+\sum_{k=1}^{K} \beta_{k}} \\
& = \underbrace{\frac{\beta_0}{\beta_0 + \sum_{k=1}^K \beta_k} \mathbf{1}\big(\mathcal{D}(\mathbb{P}_0||\mathbb{P}_1) - 2\epsilon^2 \big)}_{\text{Covert penalty}}  +
\underbrace{\frac{1}{\beta_0 + \sum_{k=1}^K \beta_k} \sum_{k=1}^K \beta_k \mathbf{1}\big(R_k^0 - R_k^{tot}\big)}_{\text{QoS penalty}},
\end{aligned}
\label{eq:penalty}
\end{equation}
where $\beta_0$ and $\beta_k$ are control variables. $\mathbf{1}(a - b)$ is the indicator function in which $\mathbf{1}(a - b) = 1$ if $a - b > 0$, and otherwise $\mathbf{1}(a - b) = 0$. The meaning of $p_t$ can be expressed as follows. 
The penalty is increased with each of covert or QoS constraints, i.e., (\ref{eq:max-min-rate}e) and (\ref{eq:max-min-rate}f), multiplying with corresponding weights $\beta_0$ and $\beta_k$ ($k = 1, 2, \ldots, K$). The penalty is then normalized so that $p_t \in [0, 1]$ (i.e., the first line of (\ref{eq:penalty})). The penalty can be rewritten as the sum of two components, i.e., covert penalty and QoS penalty as shown in the second line of (\ref{eq:penalty}).
Note that the penalty of the BS can be calculated by using the feedback mechanism. Once the users receive the messages from the BS, they calculate the data rates of the messages and send the calculated data rates back to the BS along with their minimum rate requirements (QoS requirements)~\cite{hyang2020}. Based on the feedback from users, the BS can calculate the corresponding QoS penalty. Similarly, the BS can be notified by the users if the channel is interrupted by the warden and the covert penalty can be calculated accordingly.
Our designed immediate reward aims to encourage the BS to minimize the penalty $p_t$ to 0. Thus, the max-min fairness is guaranteed while covert and QoS constraints are satisfied.

\subsection{Optimization Formulation}
\label{subsec:optimization-formulate}
We consider a stochastic policy $\Omega_{\theta}$ of the BS (i.e., $\Omega_{\theta}: \mathcal{S} \times \mathcal{A} \rightarrow [0, 1]$), as a probability that action $a_t$ is taken given the current state $s_t$, i.e., $\Omega_{\theta} = \text{Pr}\{a_t|s_t;\theta\}$, where $\theta$ is the policy parameter vector of the deep neural network. Let $J(\Omega_{\theta})$ denote the expected discounted reward of the BS by following policy $\Omega_{\theta}$:
\begin{equation}
J(\Omega_{\theta}) = \mathbb{E}_{a_t \sim \Omega, s_t \sim \mathcal{P}} \Big[\sum_{t=0}^{\infty}\tau^t r_t(s_t, a_t)\Big],
\label{eq:discounted-reward}
\end{equation}
where $\mathcal{P}(s_{t+1}|s_t, a_t)$ is the state transition probability distribution which models the dynamics of the environment, i.e., the dynamics of channel state information. Here, $\mathcal{P}$ is unknown to the BS. Our goal is to find the optimal policy $\Omega_{\theta}^*$ for the BS that maximizes $J(\Omega_{\theta}$), i.e., 
\begin{equation}
\label{eq:max-return}
\begin{aligned}
\max_{\Omega_{\theta}} \quad & J(\Omega_{\theta})\\
\textrm{s.t.} \quad & a_t \sim \Omega_{\theta}(a_t|s_t),  \\ 
\quad & s_{t+1} \sim \mathcal{P}(s_{t+1}|s_t, a_t).
\end{aligned}
\end{equation}

Maximizing $J(\Omega_{\theta})$ is very challenging as we consider that the state and action spaces are continuous. It is noted that in (\ref{eq:max-return}), we do not require the complete information of the channel, i.e., channel matrix $\mathbf{H}$ in (\ref{eq:max-min-rate}), as other works in literature~\cite{joudeh2016, joudeh2016-2, piovano2016, joudeh2016-3}. Instead, the patterns of the channel can be learned through feedback from the users with deep neural networks.
For this, we develop a learning algorithm based on a policy gradient method, namely Proximal Policy Optimization (PPO) ~\cite{schulman2017}, to approximate the optimal policy of the BS. PPO is a sample-efficient algorithm which can work under the large continuous state and action spaces and can deal with the uncertainty of the channel state. 

\section{Proximal Policy Optimization Algorithm}
\label{sec:ppo-algo}
\subsection{PPO Algorithm}
As we discussed in Section~\ref{subsec:drl-intro}, action-value methods and policy gradient methods have their own advantages and drawbacks. In action-value methods, each action of the agent can be categorized by a real positive value, e.g., Q-value~\cite{mnih2015}, and once the optimal policy is obtained, the optimal actions can be obtained by selecting the maximum action-value at each state. This family of algorithms is well studied for environments with discrete action space, e.g., a game requires a player to turn left, right, or jump. Thus, the action-value methods are suitable for discrete action space and the optimal policy can be effectively estimated if the number of actions are relatively small. However, in many cases, the action of an agent cannot be categorized by discrete action-values, e.g., a task requires controlling a robot arm by using continuous force. For this, policy gradient methods can be applied by directly estimating the policy of the agent instead of using a greedy selection over action-values. The policy of the agent can be a distribution, e.g., Gaussian, over actions. Therefore, instead of finding the action-values of the agent, policy gradient algorithms aim to find the ``shape" of the action distribution, i.e., mean and variance of the distribution. 

\begin{figure}[t]
\centering
\includegraphics[width=0.9\linewidth]{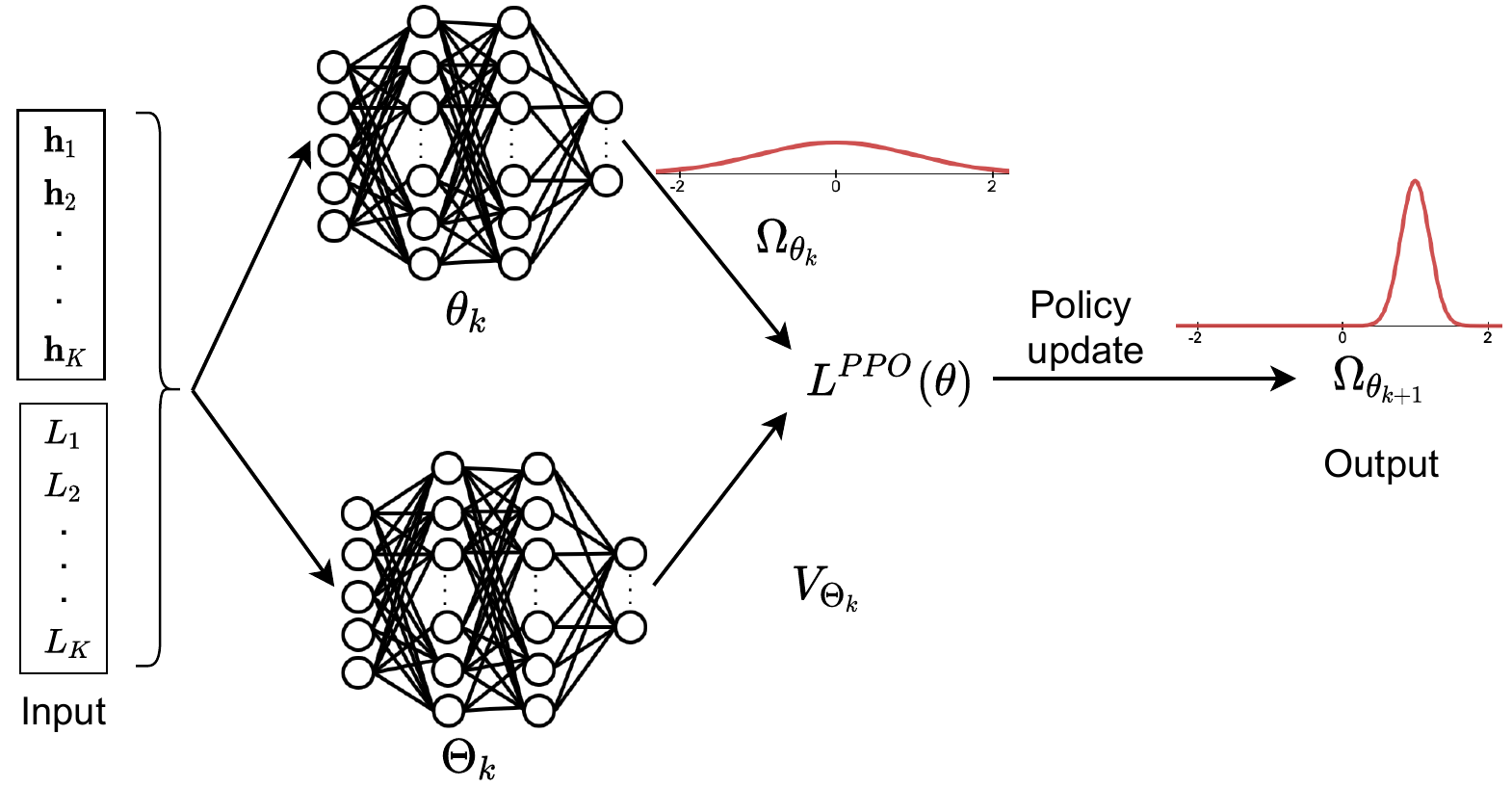}
\caption{PPO policy update at the BS.}
\label{fig:ppo-update}
\end{figure}

In our problem, all considered actions in (\ref{eq:max-min-rate}) are continuous, and thus only gradient policy methods can be used. In the following, we describe an effective algorithm based on PPO~\cite{schulman2017} to maximize the min-rate of the BS. The operation of PPO in our proposed framework is illustrated in Fig.~\ref{fig:ppo-update}. In particular, input of the policy update procedure is the joint state of channel and packets' lengths to be sent at the BS. Output is the BS's policy, i.e., action distributions. We use one Gaussian distribution to illustrate the output of the policy update in Fig.~\ref{fig:ppo-update} for the sake of presentation simplicity. In our actual implementation, the action of the BS has multiple dimensions and each dimension can be represented by a Gaussian distribution which differs in mean and variance values. The details of PPO algorithm are as follows.

The PPO uses two deep neural networks as a policy parameter vector and a value function vector, denoted by $\theta$ and $\Theta$, respectively, to efficiently update the policy. The policy parameter vector $\theta$ can be updated by using a gradient ascent method as follows:
\begin{equation}
\label{eq:theta-update}
\theta_{t+1} = \theta_t + \alpha \hat{g}_t,
\end{equation}
where $\alpha$ is the step size, and $\hat{g}_t$ is a gradient estimator. The gradient estimator $\hat{g}_t$ can be calculated by differentiating a loss function as follows:
\begin{equation}
\label{eq:gradient-estimator}
\hat{g}_t = \nabla_{\theta} L(\theta).
\end{equation}
It can be observed from (\ref{eq:theta-update}) and (\ref{eq:gradient-estimator}) that the choice of the loss function $L(\theta)$ has a significant impact on the policy update. $L(\theta)$ should have a small variance so that it does not cause bad gradient updates which result in significant decreases of $J(\Omega_{\theta})$. Since continuous action space is sensitive to the policy update, a minor negative change in updating $\theta$ can lead to destructively large policy updates~\cite{schulman2017}.
To overcome this problem, PPO algorithm uses a loss function $L^{PPO}(\theta)$ to replace $L(\theta)$:
\begin{equation}
\label{eq:loss-ppo}
 L^{PPO}(\theta) = \min \Big(\frac{\Omega_{\theta}}{\Omega_{\theta_{old}}} \hat{A}_t, u(\varepsilon, \hat{A}_t) \Big),
\end{equation}
where $\hat{A}_t$ is the advantage function and $u(\varepsilon, \hat{A}_t)$ is the clip function. $\hat{A}_t$ estimates whether the action taken is better than the policy's default behavior and $u(\cdot)$ limits significant updates which may degrade $J(\Omega_{\theta})$. The advantage function at time step $t$ can be defined by:
\begin{equation}
\hat{A}_t(s_t, a_t;\theta) = Q_t(s_t, a_t;\theta) - V_t(s_t;\Theta),
\label{eq:advantage-function}
\end{equation}
where $Q_t(s_t, a_t;\theta) = \mathbb{E}_{a_t \sim \Omega_{\theta}, s_t \sim \mathcal{P}}\Big[\sum_{l=0}^{\infty}\tau^l r_t(s_{t+l}, a_{t+l})\Big]$ is the action value function and $V_t(s_t;\Theta) = \mathbb{E}_{s_t \sim \mathcal{P}}\Big[\sum_{l=0}^{\infty}\tau^l r_t(s_{t+l}, a_{t+l})\Big]$ is the state value function. 
The clip function is thus defined as follows~\cite{schulman2017}:
\begin{equation}
u(\varepsilon, \hat{A}_t) = 
\begin{cases}
      (1+\varepsilon)\hat{A}_t, & \text{if}\ \hat{A}_t \geq 0, \\
      (1-\varepsilon)\hat{A}_t, & \text{if}\ \hat{A}_t < 0.
    \end{cases}
    \label{eq:clip-function}
\end{equation}

The idea of PPO is to prevent the new policy from being attracted to go far away from the old policy $\Omega_{old}$.  The first term inside the $\min$ operator in (\ref{eq:loss-ppo}), i.e., $\frac{\Omega_{\theta}}{\Omega_{\theta_{old}}} \hat{A}_t$, is the surrogate objective which takes into consideration the probability ratio between the new policy and old policy, i.e., $\frac{\Omega_{\theta}}{\Omega_{\theta_{old}}}$. The second term, i.e., $u(\varepsilon, \hat{A}_t)$, removes the incentive for moving this probability ratio outside of the interval $[1-\varepsilon, 1+\varepsilon]$. The pseudo-code of the PPO algorithm is described in Algorithm 1.

The main steps of the PPO algorithm can be described as follows. First, a policy parameter vector $\theta_0$ and a value function parameter vector $\Theta_0$ are randomly initialized (i.e., lines 2 and 3 in Algorithm 1). Second, at each policy update episode, numbered by $k$, the BS collects a set of trajectories $\mathcal{B}_k$, i.e., a batch of state, action, and reward values, by running current policy $\Omega_{\theta_k}$ over $T$ time steps (i.e., line 5). After that, the cumulative reward is calculated as in line 6. Next, the BS computes the advantage function as in (\ref{eq:advantage-function}) (i.e., line 6). With the obtained advantage function, the loss function can be calculated as (\ref{eq:loss-ppo}) and the policy parameter vector can be updated as line 8 in Algorithm 1. Finally, the value function parameter vector can be updated as in line 9. The procedure is repeated until the cumulative reward values converge to saturating values.

\begin{algorithm}[t]
\caption{Proximal Policy Optimization (PPO)}
 \textbf{Input}: \\
 Initialize policy parameter vector $\theta_{0}$, \\ 
 Initialize value function parameter vector $\Theta_{0}$,  \\
 \For{$k = 0, 1, 2, \ldots$}{
  Collect set of trajectories $\mathcal{B}_{k}=\left\{\upsilon_{t};\upsilon_t = (s_t, a_t, r_t)\right\}$ by running policy $\Omega_{\theta_k}$ in the environment \\
  Compute cumulative reward $\hat{R}_{t} = \sum_{t=0}^{T} \tau^t r_t$ \\
  Compute advantage function $\hat{A}_{t}$ as in (\ref{eq:advantage-function}) \\
  Update the policy by maximizing the objective (\ref{eq:loss-ppo}): 
   \begin{equation*}
\theta_{k+1}=\argmax_{\theta} \frac{1}{\left|\mathcal{B}_{k}\right| T} \sum_{\upsilon \in \mathcal{B}_{k}} \sum_{t=0}^{T} L^{PPO}(\theta),
 \end{equation*} \\
  Fit value function by regression on mean-squared error: 
  \begin{equation*}
\Theta_{k+1}=\argmin_{\Theta} \frac{1}{\left|\mathcal{B}_{k}\right| T} \sum_{\upsilon \in \mathcal{B}_{k}} \sum_{t=0}^{T}\left(V_t\left(s_{t};\Theta\right)-\hat{R}_{t}\right)^{2}
 \end{equation*}
 }
 \textbf{Outputs}: $\Omega_{\theta}^* = \text{Pr}(a_t|s_t;\theta)$
\label{algo:ppo}
\end{algorithm}

\subsection{Complexity Analysis}
We further analyze the computational complexity of the PPO algorithm used in our considered system. Since the PPO uses deep neural networks as an approximator function, the complexity mostly depends on updating these networks. As the two deep neural networks in PPO share the same architecture, the complexity of updating these networks can be analyzed as follows.
Each network consists of an input layer $X_0$, two fully-connected layers $X_1$ and $X_2$, and an output layer $X_3$. Let $|X_i|$ be the size of the layer $X_i$, i.e., the number of neurons in layer $X_i$. The complexity of the two networks can be calculated by $2(|X_0||X_1| + |X_1||X_2| + |X_2||X_3|)$. At each episode update, a trajectory, i.e., a batch of state, action, and reward values, are sampled by running the current policy to calculate the advantage function and value function to update the network. Thus, the total complexity of the training process is $O \Big(2 T |\mathcal{B}_k|(|X_0||X_1| + |X_1||X_2| + |X_2||X_3) \Big)$, where $|\mathcal{B}_k|$ is the size of the trajectory sampled from environment. There are two main reasons that PPO is a sample-efficient algorithm. First, the size of a trajectory $\mathcal{B}_k$ is relatively small, i.e., from hundreds to thousands~\cite{schulman2017}, compared with the size of a replay memory in conventional action-value methods, e.g., from 50,000 to 1,000,000 in DQN~\cite{mnih2015} . Here, in our simulations, $|\mathcal{B}_k|$ is set at 200. Second, the size of the output layer of PPO is equal to the number of action dimensions. As a result, the size of the output layer can be significantly smaller than those of action-value methods that discrete continuous action space into different chunks to compute the action values, e.g., Q-values. 
Clearly, the architecture of the deep neural networks are simple enough to be implemented in the base stations which are usually equipped with sufficient computing resources.

\section{Performance Evaluation}
\label{sec:performance-evaluation}
\subsection{Parameter Setting}
We consider our simulation parameters as follows. We use the same parameters for the RSMA and covert communications as those in~\cite{mao2018, yan2017}. The total transmission power of the BS is set to be $P_t = 20$ (dB). The control variables of channel in (\ref{eq:channel-realization}) are set at $(g_1, g_2, g_3) = (1.0, 0.8, 0.2)$, and $(\phi_1, \phi_2, \phi_3) = (0, \frac{\pi}{9}, \frac{2\pi}{9})$. Furthermore, the channel estimation errors are set at $\sigma_1^2 = P_t^{-0.6}$, $\sigma_2^2 = 0.8 P_t^{-0.6}$, and $\sigma_3^2 = 0.2 P_t^{-0.6}$~\cite{mao2018}. Those equivalent values for the warden in (\ref{eq:channel-warden}) are set as $g_w = 0.4$ and $\phi_w = \frac{\pi}{6}$.
The covert requirement parameter $\epsilon = 0.1$~\cite{yan2017}. Unlike conventional transmission schemes, covert communications require a 	relatively low data rate to hide information from the warden/adversary. Therefore, we set the QoS requirements of the users to be $R_1^0 = R_2^0 = R_3^0 = 10^{-4}$ (bps/Hz). We assume that the length of message $W_k$ to be sent at the BS follows uniform distribution with the minimum and maximum values are 0 and 1 kilobits, respectively, i.e., $L_k \sim \mathcal{U}(0, 1.0)$ (kilobits). The number of antennas at the BS and the number of users are set as $M = K = 3$.

For the deep neural networks, our parameters are set as follows. The two deep neural networks representing the policy parameter vector and the value function vector, i.e., $\theta$ and $\Theta$, respectively, share the same architecture. Each deep neural network has two fully connected layer and each layer contains 64 neurons. The number of neurons in the output layer is equal to the number of dimensions of action, i.e., $3K + 1$. The number of neurons in the input layer is equal to the number of dimensions of the joint state at the BS (as shown in Fig.~\ref{fig:ppo-update}), i.e., $2K$. The learning rate and clip values of PPO are adopted from~\cite{schulman2017}. 

In the following, we investigate the performance of our proposed PPO algorithm on RSMA and SDMA systems, denoted as P-RSMA and P-SDMA, respectively. To further understand the impacts of covert communications on both RSMA and SDMA systems, we run various simulations for scenarios in the FBL regime and infinite blocklength (IBL) regime. It is noted that in the IBL regime, the BS can achieve full data rate with Shannon capacity and the covert constraint, i.e., constraint (\ref{eq:max-min-rate}e), is temporarily removed. In the case of covert communications under consideration with FBL, the optimization problem is fully considered as in (\ref{eq:max-min-rate}). Furthermore, we introduce other baselines in which a Greedy algorithm is applied. This is to evaluate efficiency of the proposed learning algorithm. These baselines, denoted as G-RSMA and G-SDMA, aim to obtain the maximum immediate reward at each time step, compared with all the historical reward values stored in a buffer, without considering the long-term cumulative reward. In the following, we investigate the performance of all the aforementioned schemes. The considered metrics are (i) average (covert) min-rate (or average reward) and (ii) average (covert) sum-rate.
\subsection{Simulation Results}
\subsubsection{Convergence property}
\begin{figure*}
	\centering
	\begin{subfigure}[b]{0.5\textwidth}
		\centering
		\includegraphics[scale=0.47]{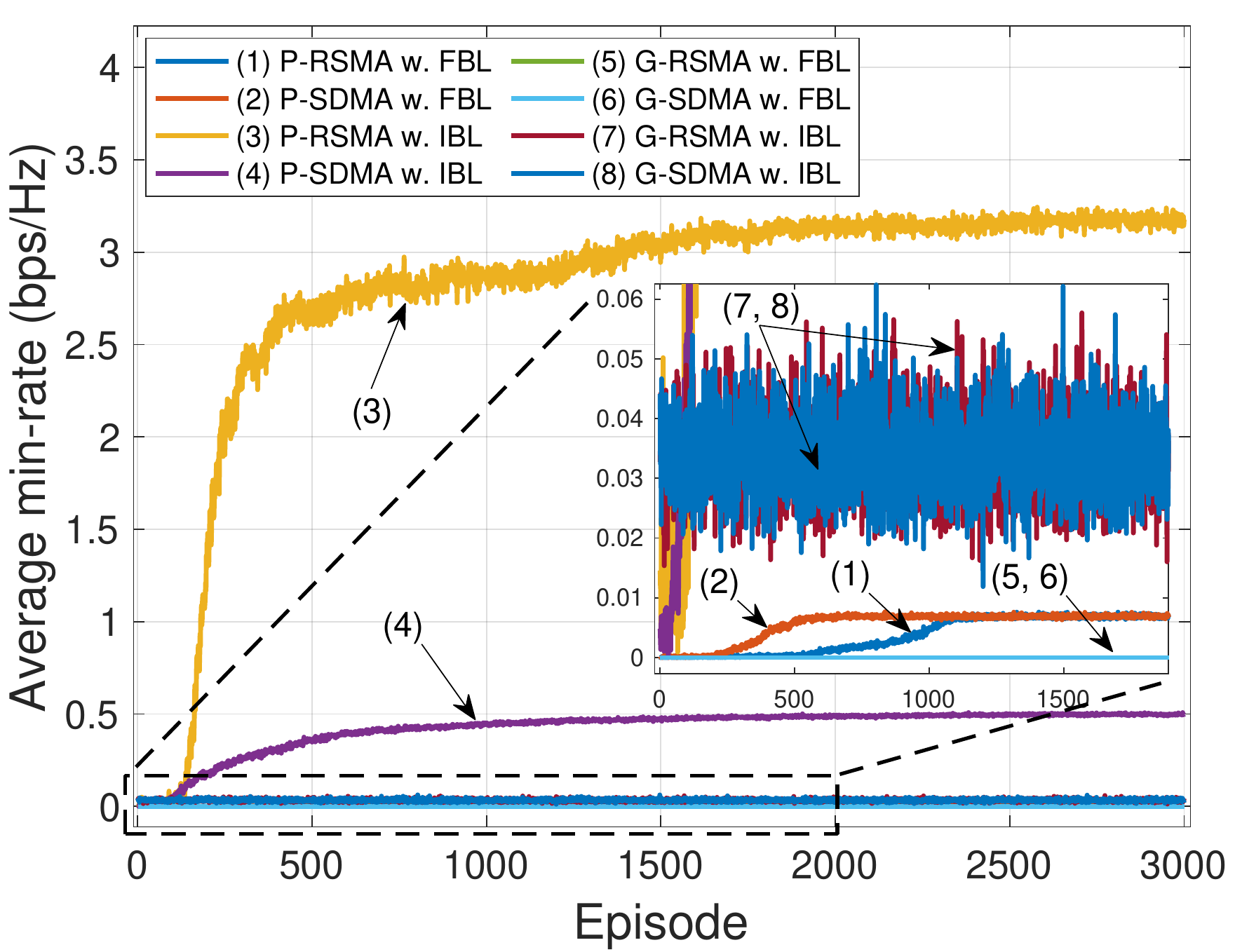}
		\caption{}
	\end{subfigure}%
	~ 
	\begin{subfigure}[b]{0.5\textwidth}
		\centering
		\includegraphics[scale=0.47]{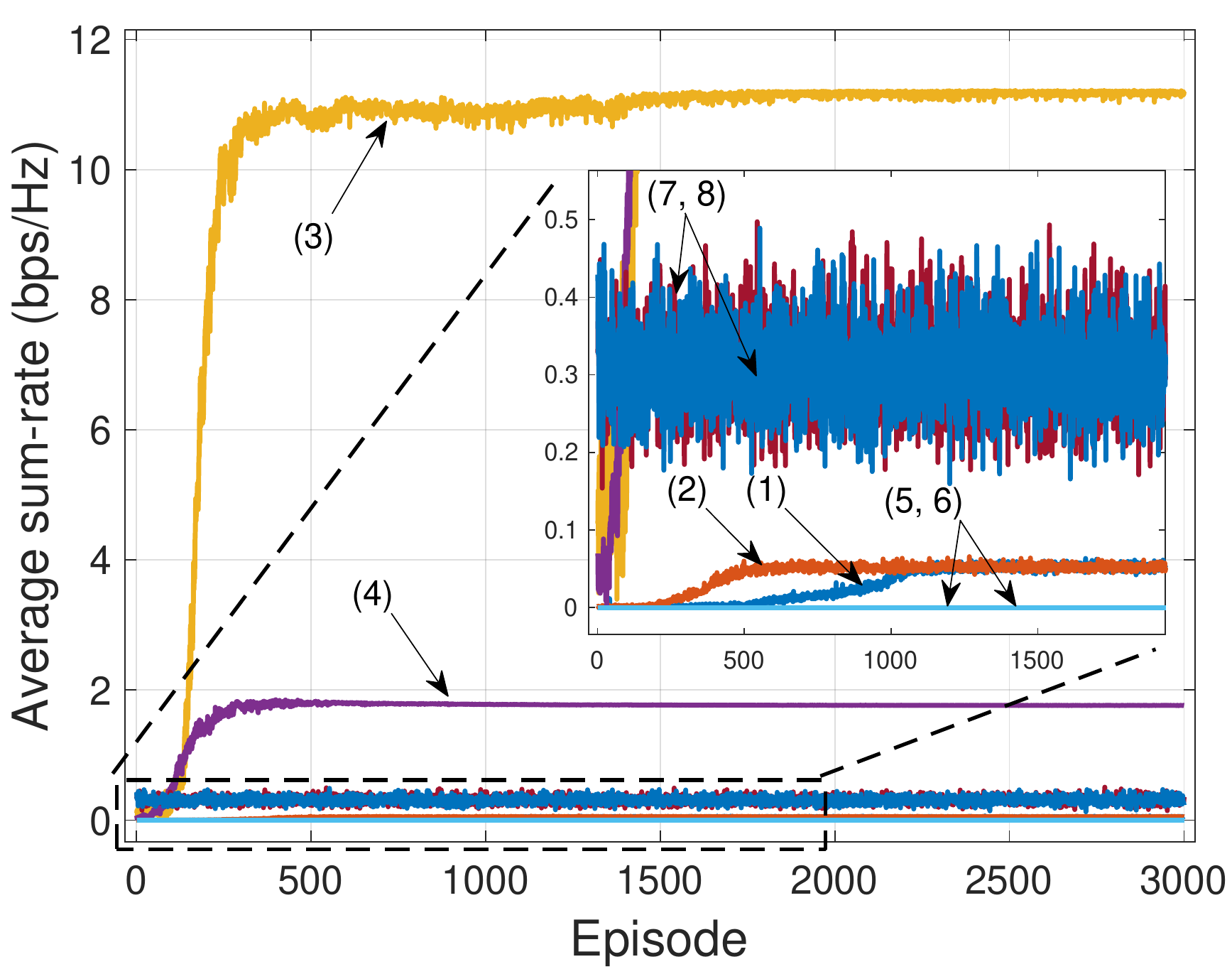}
		\caption{}
	\end{subfigure}%
	\caption{(a) Average min-rate and (b) average sum-rate with $P_t=30$ (dB) allocated for 3 users.} 
	\label{fig:convergence-curves}
\end{figure*}
We first investigate the convergence performance of the proposed P-RSMA and P-SDMA in the IBL regime (lines (3), (4), (7), and (8) in Fig.~\ref{fig:convergence-curves}). It can be observed from Fig.~\ref{fig:convergence-curves} that the proposed P-RSMA achieves the highest min-rate and sum-rate values, followed by P-SDMA (lines (3) and (4)). The reason is that in the IBL regime, the BS can obtain the data rate with full Shannon capacity. The results also suggest that RSMA outperforms SDMA in both min-rate and sum-rate, which strongly confirms that RSMA performs better than SDMA~\cite{mao2018}. In the same IBL regime, the min-rate and sum-rate obtained of G-RSMA and G-SDMA are much lower than those of P-RSMA and P-SDMA (lines (7) and (8)). The reason is that the Greedy algorithm only considers historical rewards and immediate rewards without aiming to maximize the long-term reward. Unlike the Greedy scheme, the PPO with deep neural networks can iteratively update the policy toward the maximum cumulative reward. Furthermore, bad updates negating the reward values are eliminated with clip function (\ref{eq:clip-function}). Thus, the learning curves obtained by P-RSMA and P-SDMA are much more stable.

In the FBL regime, it can be observed that the min-rate and sum-rate obtained by all the schemes are much lower than those in the IBL regime (lines (1), (2), (5), and (6)). The reason is that (i) the data rate is no longer following the Shannon capacity and (ii) the data rate is reduced close to 0 to achieve covertness~\cite{bash2013}. With G-RSMA and G-SDMA (lines (5) and (6)), the min-rate and sum-rate are 0. In other words, the Greedy algorithm can only achieve covert communications by reducing the data rate to 0 and no information can be exchanged between the BS and users. With P-RSMA and P-SDMA (lines (1) and (2)), both min-rate and sum-rate values are relatively small but remained positive when the algorithm converges. These results confirm that with the proposed PPO algorithm, the BS and users can exchange covert information without being detected by the warden.

\subsubsection{Impacts of transmission power}
\begin{figure*}
	\centering
	\begin{subfigure}[b]{0.5\textwidth}
		\centering
		\includegraphics[scale=0.47]{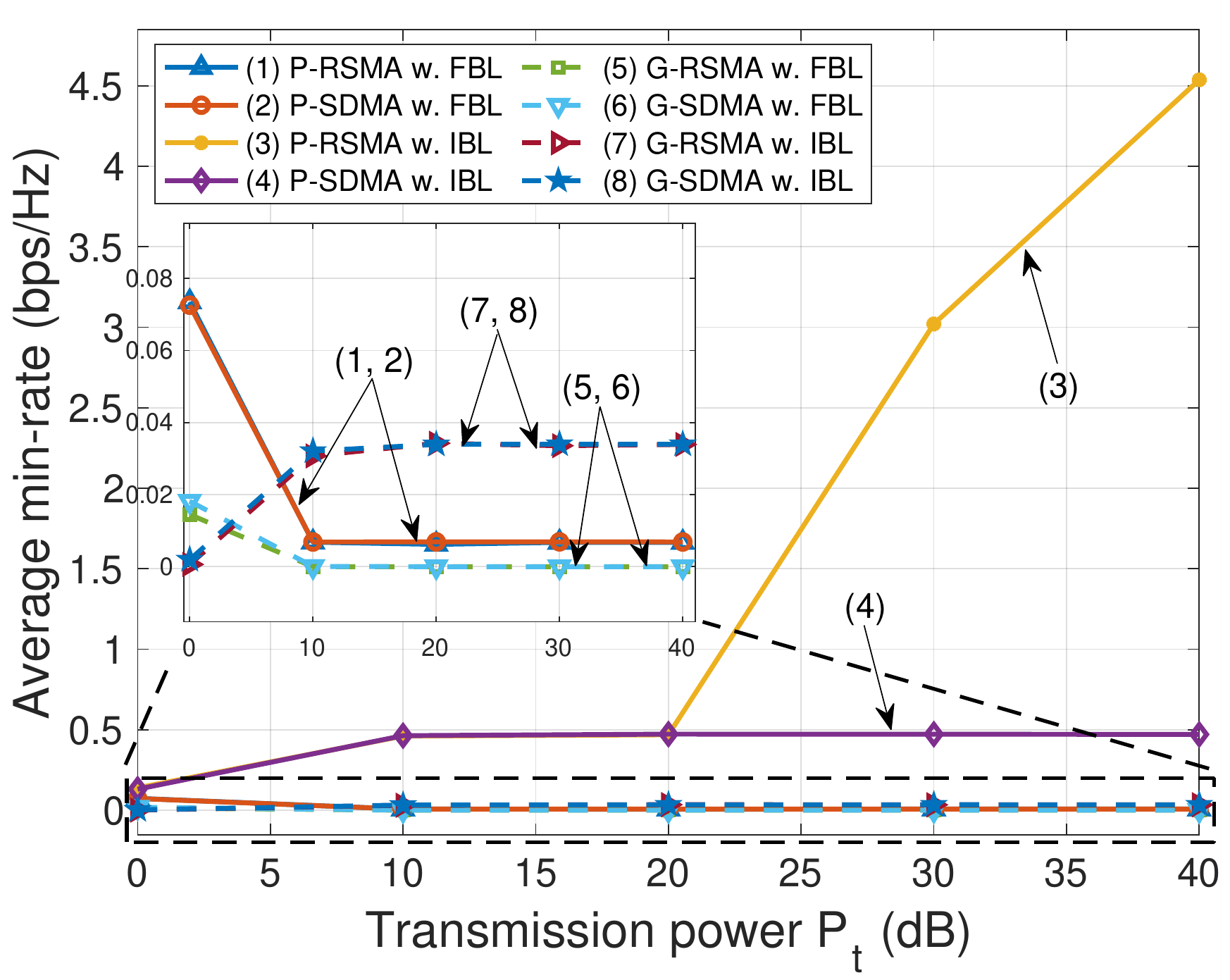}
		\caption{}
	\end{subfigure}%
	~ 
	\begin{subfigure}[b]{0.5\textwidth}
		\centering
		\includegraphics[scale=0.47]{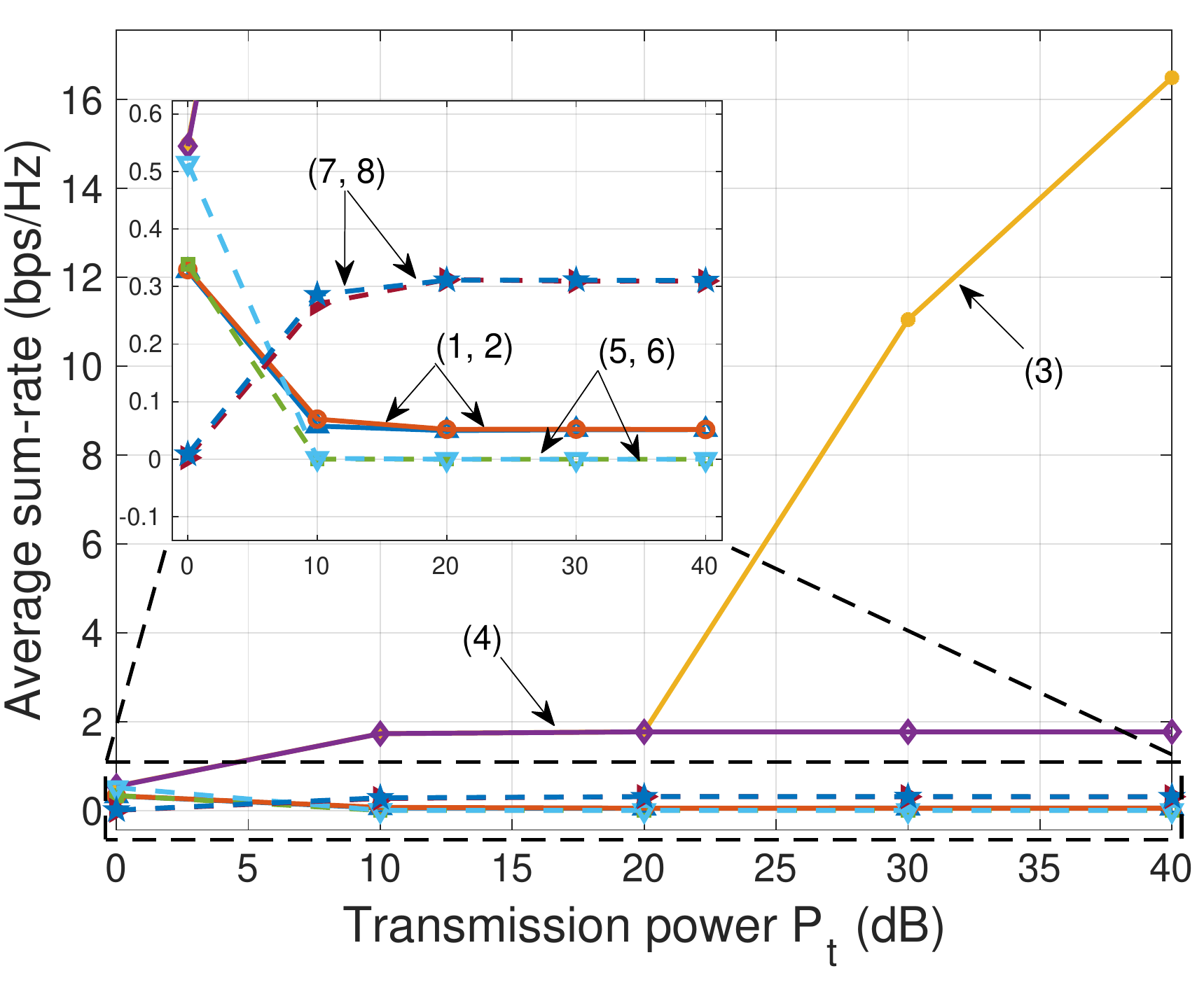}
		\caption{}
	\end{subfigure}%
	\caption{(a) Average min-rate and (b) average sum-rate vs. transmission power $P_t$ (dB).} 
	\label{fig:power-vary}
\end{figure*}

Next, in Fig.~\ref{fig:power-vary} we vary the transmission power $P_t$ at the BS and evaluate the performance of the proposed schemes. Similarly, we first discuss the performance of all the schemes in the IBL regime. It can be observed that the min-rate and sum-rate obtained by P-RSMA and P-SDMA increase with the transmission power at the BS (lines (3) and (4)). Unlike RSMA, SDMA's data rate is saturated with high transmission power~\cite{joudeh2016}. In low power transmission region (e.g., 0 dB to 20 dB), the difference between RSMA and SDMA is insignificant~\cite{dizdar2020}. Similar to the results in Fig.~\ref{fig:convergence-curves}, the min-rate and sum-rate of G-RSMA and G-SDMA are much lower than those of P-RSMA and P-SDMA. 

In the FBL regime, it can be observed that the min-rate and sum-rate obtained by P-RSMA and P-SDMA are much lower than those in the IBL regime (lines (1) and (2)). When $P_t$ increases, the data rates of P-RSMA and P-SDMA remain unchanged at 0.007 bps/Hz for the average min-rate and 0.05 bps/Hz for the average sum-rate. These results imply that with the proposed PPO algorithm, the covert communications between the BS and users can be maintained at a positive rate. In other words, the covertness can always be achieved regardless of the transmission power at the BS. Unlike the PPO, the Greedy algorithm can only hide information from the warden by reducing the data rate to 0 or no information can be exchanged (lines (5) and (6)).

\subsubsection{Impacts of covert constraint}
\begin{figure*}
	\centering
	\begin{subfigure}[b]{0.5\textwidth}
		\centering
		\includegraphics[scale=0.47]{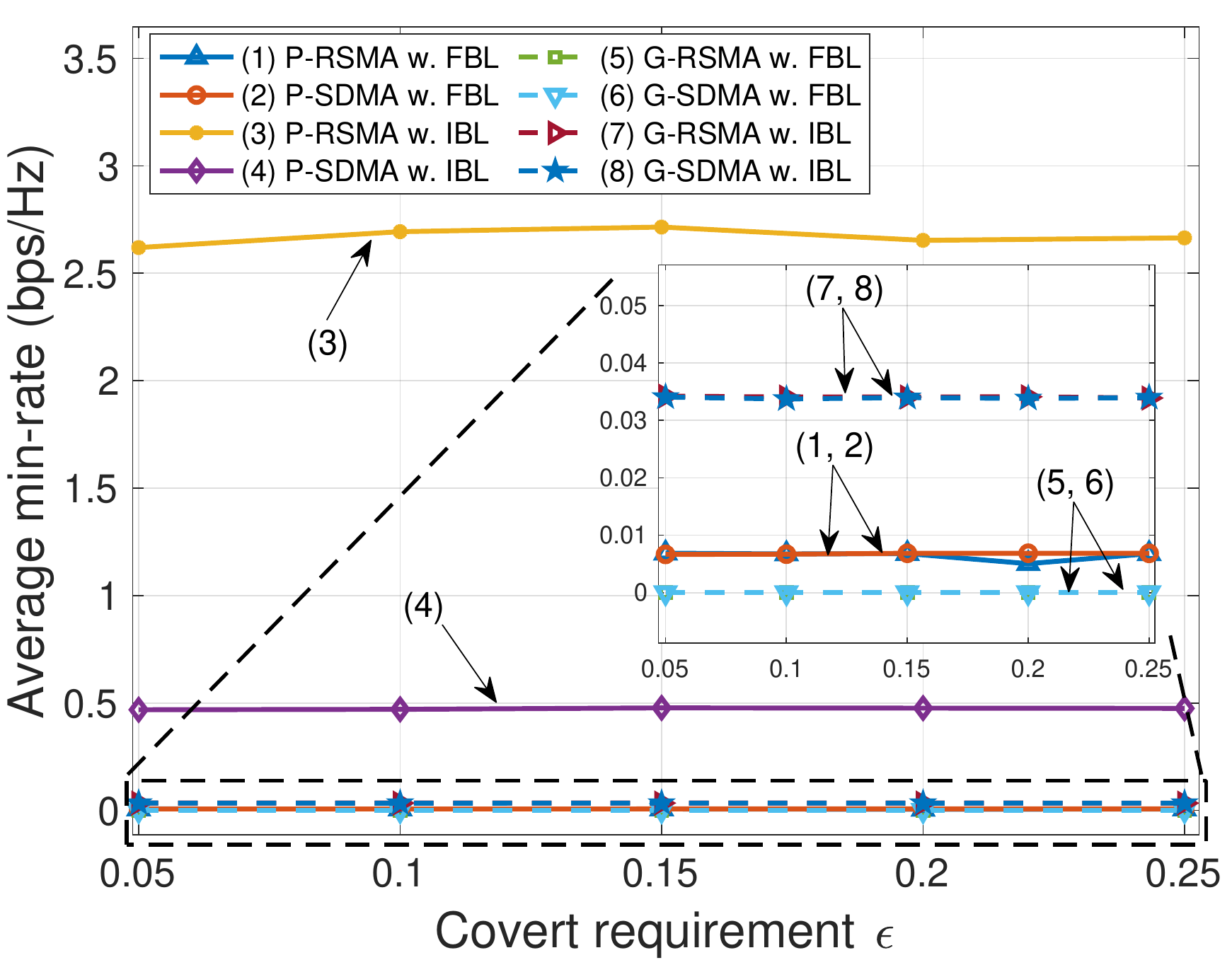}
		\caption{}
	\end{subfigure}%
	~ 
	\begin{subfigure}[b]{0.5\textwidth}
		\centering
		\includegraphics[scale=0.47]{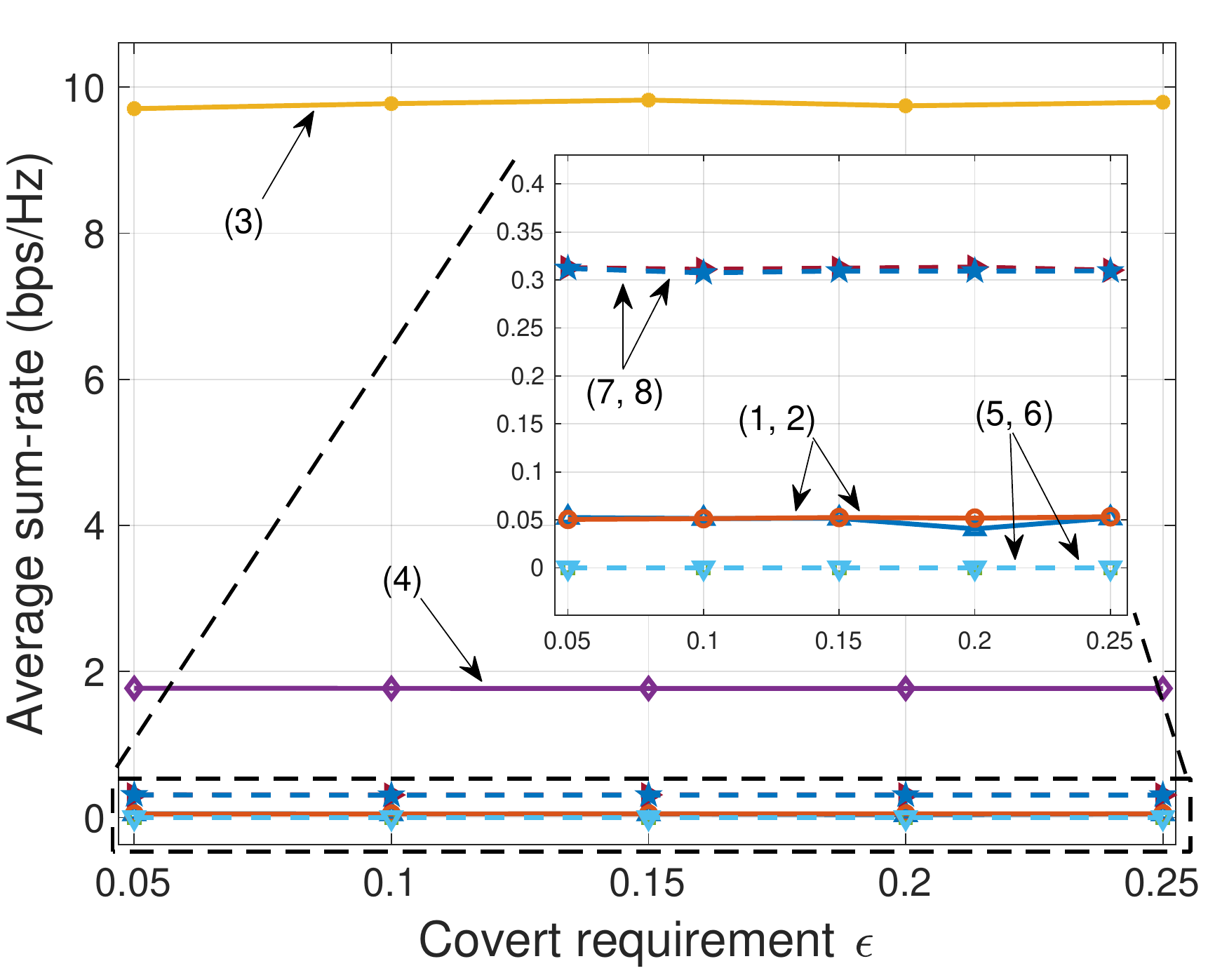}
		\caption{}
	\end{subfigure}%
	\caption{(a) Average min-rate and (b) average sum-rate vs. covert requirement ($\epsilon$).} 
	\label{fig:epsilon-vary}
\end{figure*}

In Fig.~\ref{fig:epsilon-vary}, we evaluate the impacts of the covert constraint to the system performance by varying $\epsilon$ in (\ref{eq:max-min-rate}e). Similarly, we first discuss the results in the IBL regime. Since the impacts of the covert constraint are not considered in this regime, it is clearly observed that the average min-rate and sum-rate values of P-RSMA and P-SDMA (lines (3) and (4)) remain stable and significantly higher than those of the baselines G-RSMA and G-SDMA (lines (7) and (8)) with the increase of the covert constraint parameter $\epsilon$. 

In the FBL regime, the average min-rate and sum-rate values obtained by P-RSMA and P-SDMA remain stable as $\epsilon$ increases (lines (1) and (2)). In particular, these saturated values are 0.007 bps/Hz for min-rate and 0.05 bps/Hz for sum-rate. For the baselines G-RSMA and G-SDMA (lines (5) and (6)), the obtained min-rate and sum-rate values are equal to 0, which illustrates that these baselines cannot achieve covert transmissions in the considered setting.

\subsubsection{Impacts of blocklength}
\begin{figure*}
	\centering
	\begin{subfigure}[b]{0.5\textwidth}
		\centering
		\includegraphics[scale=0.47]{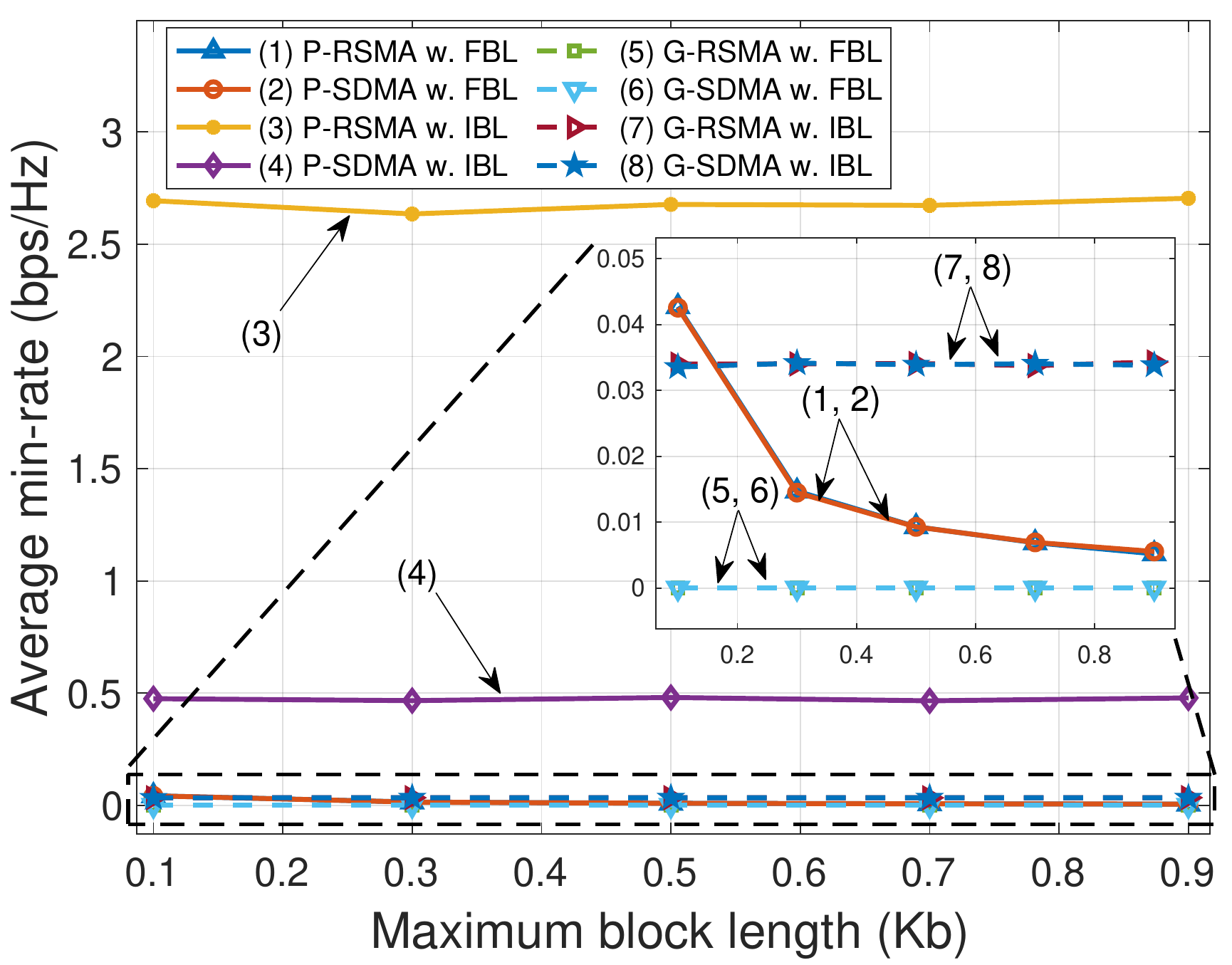}
		\caption{}
	\end{subfigure}%
	~ 
	\begin{subfigure}[b]{0.5\textwidth}
		\centering
		\includegraphics[scale=0.47]{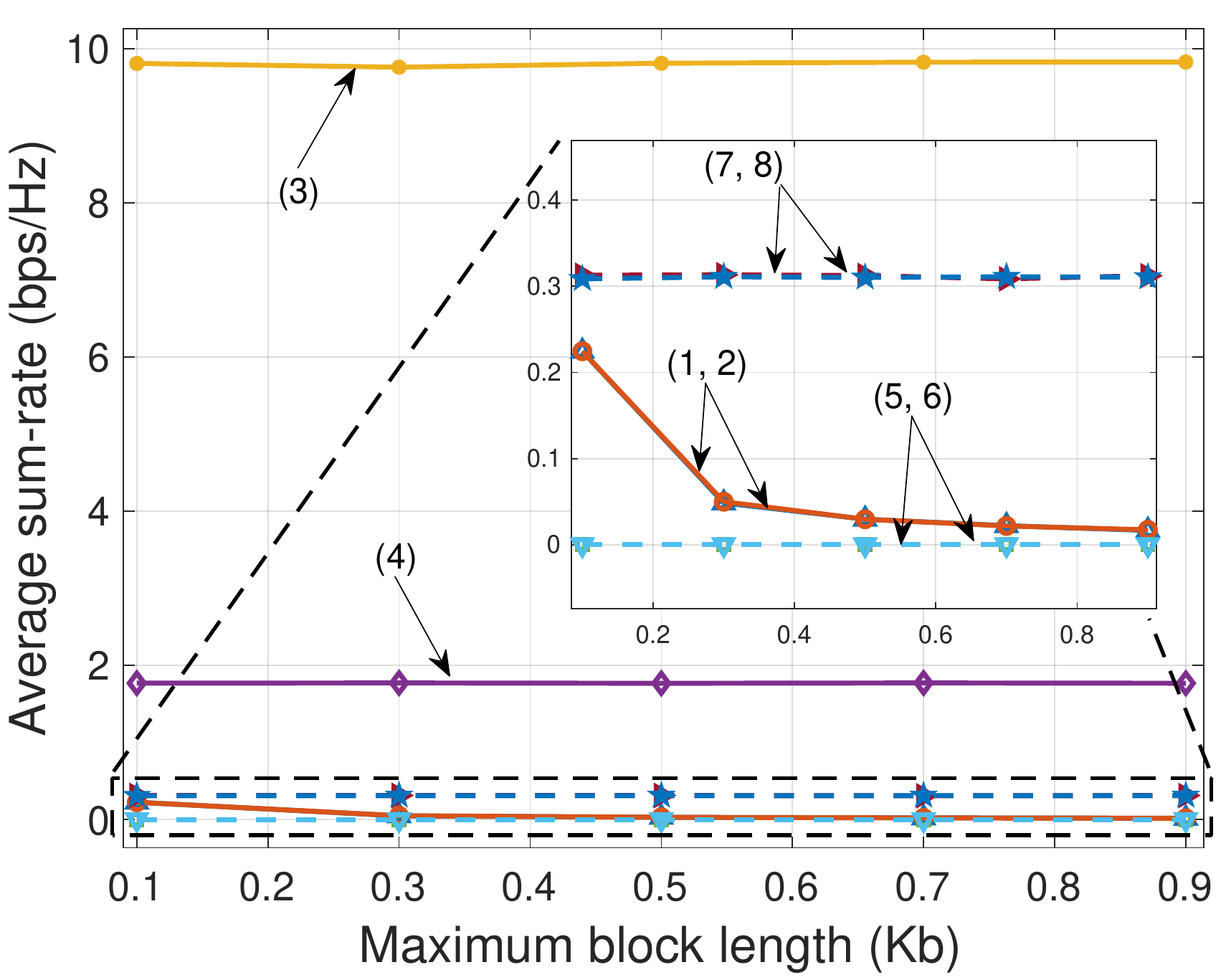}
		\caption{}
	\end{subfigure}%
	\caption{(a) Average min-rate and (b) average sum-rate vs. maximum blocklength ($L_k$).} 
	\label{fig:pcklen-vary}
\end{figure*}

Finally, in Fig.~\ref{fig:pcklen-vary}, we investigate the impacts of blocklength, i.e., the length $L_k$ of the message $W_k$ being sent at the BS, to the system performance. We vary the distributions of the packet length $L_k$ at the BS with different intervals. In particular, we consider nine distribution intervals that are $\mathcal{U}[0, 0.1], \mathcal{U}[0.1, 0.2]$, $\mathcal{U}[0.2, 0.3], \ldots, \mathcal{U}[0.8, 0.9]$ (Kilobits). Note that in Fig.~\ref{fig:pcklen-vary}, we denote these distributions by their maximum values for the sake of simplicity. It can be observed that, in the IBL regime where $L_k \rightarrow \infty$, the values of min-rate and sum-rate of all the schemes are independent with the blocklength (lines (3), (4), (7), and (8)). However, in the FBL regime, the min-rate and sum-rate values obtained by P-RSMA and P-SDMA decrease as the blocklength increases (lines (1) and (2)). In other words, the higher the blocklength of message is sent from the BS, the lower the data rate can be achieved. This finding is similar to mathematical analysis derived in~\cite{bash2013}. According to~\cite{bash2013}, the number of bits that can be covertly transmitted, denoted as $n$, asymptotically approaches zero and $n$ follows a square root law, i.e., $\mathcal{O}(\sqrt{n})/n \rightarrow 0$ as $n \rightarrow 0$. Unlike the positive data rate  values achieved by the proposed schemes, the min-rate and sum-rate values of the baselines G-RSMA and G-SDMA remain at 0 (lines (5) and (6)).

\section{Conclusion}
\label{sec:conclusion}
In this paper, we have developed a novel dynamic framework to jointly optimize power allocation and rate control for the RSMA networks under the uncertainty of surrounding environment and with requirements about covert communications. In particular, our proposed stochastic optimization framework can adjust its transmission power together with message splitting based on its observations from surrounding environment to maximize the rate performance for the whole system. Furthermore, we have developed a learning algorithm that can not only help to BS to deal with continuous action and state spaces effectively, but also quickly find the optimal policy for the BS without requiring the completed information about surrounding environment in advance. Extensive simulations have demonstrated that with the obtained policy, the BS can dynamically adjust power and transmission rates to users, so that the achievable covert rate can be maximized. At the same, the BS can minimize the probability of being detected by the warden.

\appendix
\subsection{Relative entropy $\mathcal{D}(\mathbb{P}_0 | \mathbb{P}_1)$ between two distributions $\mathbb{P}_0$ and $\mathbb{P}_1$}
\label{appe:kl}

As defined in the hypothesis test of the warden in (\ref{eq:hypothesis-test}), we have the distribution of the i.i.d. Gaussian random variables with variance $\sigma_w^2$ is $\mathbb{P}_0 = \mathcal{N}(0, \sigma_w^2)$, which corresponds to the case when the BS is not transmitting. Note that the warden does not know the codebook. Therefore, the warden's probability distribution of the transmitted symbols is of zero-mean i.i.d. Gaussian random variables with variance $P_f$. Since we have the signal $\mathbf{x}$ is transmitted with power $P_t$ at the transmitter and channel between the warden and the BS is defined in (\ref{eq:channel-warden}), we have $P_f = g_w P_t$. Therefore, the distribution of $\mathbb{P}_1$ is as follows:
\begin{equation}
\begin{split}
\mathbb{P}_1 &= \mathcal{N}(0, P_f + \sigma_w^2) \\
&= \mathcal{N}(0, g_w P_t + \sigma_w^2).
\end{split}
\end{equation} 
Let  $a = \sigma_w$ and $b = \sqrt{g_w P_t + \sigma_w^2}$, we have the respective probability distribution functions of $\mathbb{P}_0$ and $\mathbb{P}_1$ are as follows:

\begin{equation}
p_0(x) = \frac{1}{\sqrt{2 \pi} a} \mathrm{e}^{\frac{-1}{2} (\frac{x}{a})^2},
\end{equation}

\begin{equation}
p_1(x) = \frac{1}{\sqrt{2 \pi} b} \mathrm{e}^{\frac{-1}{2} (\frac{x}{b})^2}.
\end{equation}

The relative entropy between $\mathbb{P}_0$ and $\mathbb{P}_1$ is then calculated by:

\begin{equation}
\begin{split}
\mathcal{D}(\mathbb{P}_0 | \mathbb{P}_1) &= \int_{-\infty}^{+\infty} p_0(x) \ln \frac{p_0(x)}{p_1(x)} \mathrm{~d} x \\
&= \int_{-\infty}^{+\infty} \frac{1}{\sqrt{2 \pi} a} \mathrm{e}^{\frac{-1}{2} (\frac{x}{a})^2} \ln \Big(\frac{b}{a} e^{\frac{-1}{2} \big[(\frac{x}{a})^2 - (\frac{x}{b})^2 \big]} \Big) \mathrm{~d} x \\
&= \int_{-\infty}^{+\infty} \frac{1}{\sqrt{2 \pi} a} \mathrm{e}^{\frac{-1}{2} (\frac{x}{a})^2} \Big(\ln \big(\frac{b}{a}\big) - \frac{1}{2} \big[\big(\frac{x}{a}\big)^2 - \big(\frac{x}{b} \big)^2 \big] \Big) \mathrm{~d} x \\
&= -\frac{1}{2^{\frac{3}{2}} \sqrt{\pi} a^{3} b^{2}} \underbrace{\int_{-\infty}^{+\infty} \left(\left(b^{2}-a^{2}\right) x^{2}-2 a^{2} b^{2} \ln \left(\frac{b}{a}\right)\right) \mathrm{e}^{-\frac{x^{2}}{2 a^{2}}} \mathrm{~d} x}_{\mathcal{D}_1} \text{ (apply linearity)}.
\end{split}
\end{equation}

Now we need to solve $\mathcal{D}_1$. We expand $\mathcal{D}_1$ and apply linearity:

\begin{equation}
\begin{split}
\mathcal{D}_1 &= \int_{-\infty}^{+\infty}\left(\left(b^{2}-a^{2}\right) x^{2} e^{-\frac{x^{2}}{2 a^{2}}}-2 a^{2} b^{2} \ln \left(\frac{b}{a}\right) \mathrm{e}^{-\frac{x^{2}}{2 a^{2}}}\right) d x \\
& = \left(b^{2}-a^{2}\right) \underbrace{\int_{-\infty}^{+\infty} x^{2} \mathrm{e}^{-\frac{x^{2}}{2 a^{2}}} \mathrm{~d} x}_{\mathcal{D}_2} - 2 a^{2} b^{2} \ln \left(\frac{b}{a}\right) \underbrace{\int_{-\infty}^{+\infty} \mathrm{e}^{-\frac{x^{2}}{2 a^{2}}} \mathrm{~d} x}_{\mathcal{D}_3}
\end{split}
\label{eq:d1}
\end{equation}

We first solve $\mathcal{D}_2$. For this, we integrate $\mathcal{D}_1$ by parts, i.e., $\int f g' = fg - \int f'g$. Let $f = x$ and $g' = x\mathrm{e}^{-\frac{x^2}{2a^2}}$, we can calculate $f' = 1$ and $g = -a^2 \mathrm{e}^{-\frac{x^2}{2a^2}}$. We now have:

\begin{equation}
\mathcal{D}_2 = -a^{2} x \mathrm{e}^{-\frac{x^{2}}{2 a^{2}}}- \underbrace{\int_{-\infty}^{+\infty}-a^{2} \mathrm{e}^{-\frac{x^{2}}{2 a^{2}}} \mathrm{~d} x}_{\mathcal{D}_4}.
\end{equation}

$\mathcal{D}_4$ can be solved as follows. We substitute $u = \frac{x}{\sqrt{2}a}\rightarrow \mathrm{~d} x = \sqrt{2} a \mathrm{~d} u$. $\mathcal{D}_4$ becomes:

\begin{equation}
\mathcal{D}_4 = -\frac{\sqrt{\pi} a^{3}}{\sqrt{2}} \int_{-\infty}^{+\infty} \frac{2 \mathrm{e}^{-u^{2}}}{\sqrt{\pi}} \mathrm{d} u.
\end{equation}

Note that we have a special integral in $\mathcal{D}_4$, i.e., $\int_{-\infty}^{+\infty} \frac{2 \mathrm{e}^{-u^{2}}}{\sqrt{\pi}} \mathrm{d} u = \text{ erf}(u)$ is a Gauss error function. Let's plug in $\mathcal{D}_4$:

\begin{equation}
\begin{split}
\mathcal{D}_4 &= -\frac{\sqrt{\pi} a^{3} \operatorname{erf}(u)}{\sqrt{2}} \\
&= -\frac{\sqrt{\pi} a^{3} \operatorname{erf}\left(\frac{x}{\sqrt{2} a}\right)}{\sqrt{2}} \text{ (undo substitution $u = \frac{x}{\sqrt{2}a}$)}.
\end{split}
\end{equation}

Plug $\mathcal{D}_4$ in $\mathcal{D}_2$:

\begin{equation}
\mathcal{D}_2 = \frac{\sqrt{\pi} a^{3} \operatorname{erf}\left(\frac{x}{\sqrt{2} a}\right)}{\sqrt{2}}-a^{2} x \mathrm{e}^{-\frac{x^{2}}{2 a^{2}}}.
\label{eq:d2}
\end{equation}

Once $\mathcal{D}_2$ is solved, $\mathcal{D}_3 = \int_{-\infty}^{+\infty} \mathrm{e}^{-\frac{x^{2}}{2 a^{2}}} \mathrm{~d} x$ can be calculated as follows.
Let's substitute $u = \frac{x}{\sqrt{2} a} \rightarrow \mathrm{~d} x = \sqrt{2}a \mathrm{~d} u$. $\mathcal{D}_3$ becomes:

\begin{equation}
\mathcal{D}_3 = \frac{\sqrt{\pi} a}{\sqrt{2}} \int_{-\infty}^{+\infty} \frac{2 \mathrm{e}^{-u^{2}}}{\sqrt{\pi}} \mathrm{d} u.
\end{equation}

Use the previous result of Gauss error function, we have:

\begin{equation}
\begin{split}
\mathcal{D}_3 &= \frac{\sqrt{\pi} a \operatorname{erf}(u)}{\sqrt{2}} \\
&= \frac{\sqrt{\pi} a \operatorname{erf}\left(\frac{x}{\sqrt{2} a}\right)}{\sqrt{2}} \text{ (undo substitution $u = \frac{x}{\sqrt{2}a}$)}.
\end{split}
\label{eq:d3}
\end{equation}

Once $\mathcal{D}_2$ and $\mathcal{D}_3$ are solved, let's plug (\ref{eq:d2}) and (\ref{eq:d3}) into (\ref{eq:d1}):

\begin{equation}
\begin{split}
\mathcal{D}_1 &= (b^2 - a^2) \mathcal{D}_2 - 2a^2b^2 \ln\left(\frac{b}{a}\right) \mathcal{D}_3 \\
&= -\sqrt{2} \sqrt{\pi} a^{3} b^{2} \ln \left(\frac{b}{a}\right) \operatorname{erf}\left(\frac{x}{\sqrt{2} a}\right)+\frac{\sqrt{\pi} a^{3} \cdot\left(b^{2}-a^{2}\right) \operatorname{erf}\left(\frac{x}{\sqrt{2} a}\right)}{\sqrt{2}}-a^{2} \cdot\left(b^{2}-a^{2}\right) x \mathrm{e}^{-\frac{x^{2}}{2 a^{2}}}
\end{split}
\end{equation}

Finally, we have:

\begin{equation}
\begin{split}
\mathcal{D}(\mathbb{P}_0 | \mathbb{P}_1) &= -\frac{1}{2^{\frac{3}{2}} \sqrt{\pi} a^{3} b^{2}} \mathcal{D}_1 \\
&= \Eval{\left( \frac{\ln \left(\frac{b}{a}\right) \operatorname{erf}\left(\frac{x}{\sqrt{2} a}\right)}{2}-\frac{\left(b^{2}-a^{2}\right) \operatorname{erf}\left(\frac{x}{\sqrt{2} a}\right)}{4 b^{2}}+\frac{\left(b^{2}-a^{2}\right) x \mathrm{e}^{-\frac{x^{2}}{2 a^{2}}}}{2^{\frac{3}{2}} \sqrt{\pi} a b^{2}} \right)}{-\infty}{+\infty} \\
&= \ln(b) - \ln(a) + \frac{a^2}{2b^2} - \frac{1}{2}.
\end{split}
\end{equation}

Undo substitution for $a = \sigma_w$ and $b = \sqrt{g_w P_t + \sigma_w^2}$, we have:

\begin{equation}
\mathcal{D}(\mathbb{P}_0 | \mathbb{P}_1) = \ln\left(\sqrt{g_w P_t + \sigma_w^2}\right) - \ln\left(\sigma_w\right) + \frac{\sigma_w^2}{2\left(g_w P_t + \sigma_w^2\right)} - \frac{1}{2}.
\end{equation} 

The proof of Proposition~\ref{prop:kl-div} is now completed.

\end{document}